\documentclass{aa}
\input{psfig}

\def\tvi(#1,#2){\vrule height #1pt depth #2pt width 0pt}
\def\p{\partial}
\def\e{{\rm e}}
\def\d{{\rm d}}
\def\ie{i.e. }
\def\eg{e.g. }
\def\etal{et al. }
\def\rso{r_{\rm s}}
\def\cs{c_{\rm s}}
\def\M{{\cal M}}
\def\cpl{{\cal Q}_{S}}
\def\vpl{{\cal Q}_{K}}
\def\cmo{{\cal R}_{s}}

\def\eiwv{\e^{i\omega\int_R^r {\d r\over v}}}
\def\omcut{\omega_l^{\rm cut}}
\def\omcutg{\omega_{l\ge1}^{\rm cut}}
\def\omcutu{\omega_1^{\rm cut}}
\def\omcutz{\omega_0^{\rm cut}}
\def\bcmo{{\bar\cmo}}
\def\dMs{{\dot\M}_{\rm s}}
\def\cl{c_{\rm light}}

\newlength{\largeur}
\newlength{\saut}
\def\marge#1{
\setlength{\largeur}{\columnwidth}
\addtolength{\largeur}{-#1}
\setlength{\saut}{0.5\largeur}\hspace*{\saut}} \def\picture #1 by #2
(#3){
\marge{#1} \vbox to #2{
\hrule width #1 height 0pt depth 0pt
\vfill
\special{picture #3}}}
\def\scaledpicture #1 by #2 (#3 scaled #4){{
\dimen0=#1 \dimen1=#2
\divide\dimen0 by 1000 \multiply\dimen0 by #4 \divide\dimen1 by 1000
\multiply\dimen1 by #4 \picture \dimen0 by \dimen1 (#3 scaled #4)}}

\begin{document}

\thesaurus{06
(02.01.2;
02.08.1;
02.09.1;
02.19.1;
08.02.1;
13.25.5)}

\title{Entropic-acoustic instability of shocked Bondi accretion
I. What does perturbed Bondi accretion sound like ?}
\author{T. Foglizzo\thanks{e-mail: {\tt foglizzo@cea.fr}} 
}

\institute {Service d'Astrophysique, CEA/DSM/DAPNIA, CE-Saclay, 91191
Gif-sur-Yvette, France 
}

\date{Received 17 October 2000 / Accepted 22 December 2000}
\titlerunning{What does perturbed Bondi accretion sound like ?}
\maketitle

\begin{abstract}
In the radial flow of gas into a black hole (\ie Bondi accretion), the
infall of any entropy or vorticity perturbation produces acoustic waves
propagating outward. The dependence of this acoustic flux on the shape of
the perturbation is investigated in detail. This is the key process in the
mechanism of the entropic-acoustic instability proposed by Foglizzo\& Tagger
(2000) to explain the instability of Bondi-Hoyle-Lyttleton accretion. These
acoustic waves create new entropy and vorticity perturbations when they
reach the shock, thus closing the entropic-acoustic cycle. With an adiabatic
index $1<\gamma\le 5/3$, the linearized equations describing the perturbations
of the Bondi flow are studied analytically and solved numerically. The
fundamental frequency of this problem is the cut-off frequency of acoustic
refraction, below which ingoing acoustic waves are refracted out. This
cut-off is significantly smaller than the Keplerian
frequency at the sonic radius and depends on the latitudinal number $l$ of
the perturbations.When advected adiabatically inward, entropy and
vorticity perturbations trigger acoustic waves propagating outward,
with an efficiency which is highest for non radial perturbations $l=1$.
The outgoing acoustic flux produced by
the advection of vorticity perturbations is always moderate and peaks at
rather low frequency. By contrast, the acoustic flux
produced by an entropy wave is highest close to the refraction cut-off.
It can be very large if $\gamma$ is close to $5/3$.
These results suggest that the shocked Bondi flow with $\gamma=5/3$ is 
strongly unstable with respect to the entropic-acoustic mechanism.


\end{abstract}

\section{Introduction}

The time variability of the emission from X-ray binaries may be
related to hydrodynamic instabilities in the process of accretion onto
a compact star. The simplest accretion flow is the
stationary, spherically symmetric accretion of an adiabatic gas onto a
compact object, first studied by Bondi (1952).
The stability of Bondi accretion was established by Garlick (1979)
and Petterson \etal (1980) in the newtonian limit, and by Moncrief (1980) 
in the framework of general relativity. Kovalenko \& Eremin~(1998)
suggested that non radial perturbations might reach non linear amplitudes in
the supersonic part of the flow if the size of the accretor is small enough.
This effect was invisible in the 3-D numerical simulations of the Bondi flow
by Ruffert (1994a), which confirmed its stability. By contrast, a strong
instability appeared in the numerical simulations of Bondi-Hoyle-Lyttleton
accretion when a bow shock is present (\eg Ruffert 1994b for 
$\gamma=5/3$). The Bondi-Hoyle-Lyttleton accretion is named after the 
pioneering works of Hoyle \& Lyttleton (1939) and Bondi \& Hoyle (1944)
concerning an accretor moving
at supersonic speed with respect to the gas. Foglizzo \& Tagger (2000)
(hereafter FT2000) suggested that this instability might be explained 
by what they called the entropic-acoustic instability. This generic 
mechanism of instability in shocked converging flows is based on the cycle 
of entropic and acoustic waves in the subsonic region of
the flow between the shock and the sonic surface surrounding the accretor.
The advection of entropy perturbations towards the accretor produces
outgoing acoustic waves which propagate towards the shock. Perturbed
by these acoustic waves, the shock produces new entropy perturbations, thus
closing the entropic-acoustic cycle. According to FT2000, this cycle is
unstable (\ie the amplitude of the new entropy perturbation exceeds 
the initial one) if the sound speed at the sonic point is much larger than 
the sound speed at the shock. This mechanism might play a destabilizing 
role in various
astrophysical environments such as wind accretion or disc accretion.
In this series of two papers, we describe the entropic-acoustic instability
in the Bondi flow, including a spherical shock as an outer boundary (as in
Foglizzo \& Ruffert 1997). The study of this rather academic configuration
serves two purposes:
\par (i) it is the first detailed description of the entropic-acoustic
instability in a specific flow, beyond the general concepts and
approximations of FT2000.
The Bondi flow is simple enough to allow for an analytical treatment of 3D
perturbations, including non radial perturbations of entropy and
perturbations of vorticity.
\par (ii) understanding the instability of the Bondi flow with a spherical
shock might be the first step towards a comprehension of the instability of
non radial shocked accretion flows such as supersonic Bondi-Hoyle-Lyttleton
accretion.\\
In this first paper, we investigate the key part of the entropic-acoustic
instability: the excitation of outgoing acoustic waves by the advection of
entropy and vorticity perturbations. As explained in FT2000, the efficiency
of this process is directly linked to the efficiency of the refraction of
acoustic waves, which is also studied. The efficiencies of these processes
depend a priori on the frequency $\omega$ of the perturbation, its
wavenumbers in spherical coordinates $(l,m)$, and the adiabatic index
$1<\gamma\le5/3$ of the gas. The accretor is a compact object such as a
black hole, absorbing all the matter falling inside the sonic radius. The
shock, which plays no role in this first paper,  is introduced in the second
paper of this series (Foglizzo~2001) in order to close the entropic-acoustic
cycle and enable a description of the eigenmodes of the global
instability.\\
The present paper is organized as follows. Linearized equations are
established in Sect.~2. The refraction of acoustic waves is studied in
Sect.~3. The excitation of acoustic waves by the advection of entropy and
vorticity perturbations is examined in Sect.~4.

\section{Linear perturbations of pressure, entropy and vorticity in the 
Bondi flow}

\subsection{Properties of the unperturbed flow}

We consider the radial accretion flow of a gas with constant sound 
velocity $c_\infty$ and density $\rho_{\infty}$ at 
infinity, accelerated towards a point-like accretor of mass $M$, 
which is totally absorbing. The 
equations describing the unperturbed Bondi flow are recalled in 
Appendix~A. In what follows, densities and velocities are normalized by 
$c_\infty$ and $\rho_{\infty}$, and distances are normalized to the Bondi 
radius $GM/c_\infty^2$. After this normalization, the unperturbed flow 
depends on a single parameter $\gamma$, the adiabatic index of the gas.
The sound velocity $\cs$ at the sonic radius $\rso$ in the 
Bondi flow diverges for $\gamma=5/3$:
\begin{eqnarray}
\rso&=&{5-3\gamma\over4},\label{rson}\\
\cs&=&\left({2\over5-3\gamma}\right)^{1\over2}\label{cson}.
\end{eqnarray}
The "natural" sonic frequency $\cs/\rso$ is directly related to the 
Keplerian frequency $\omega_{K}$ at the sonic radius:
\begin{equation}
\omega_{K}=2^{1\over2}{\cs\over\rso}.
\end{equation}

\subsection{Differential equation for small perturbations}

The Euler equation, the equations of conservation of entropy and mass 
in the Bondi flow are linearized for small perturbations in 
Appendix~B. After a Fourier transform in time, the perturbations are 
projected onto a basis of spherical harmonics $Y_l^m$, where $l,m$ are
the numbers associated to the latitudinal ($\theta$) and longitudinal 
($\varphi$) angles. Kovalenko \& Eremin (1998) studied the same 
set of equations in order to analyze the asymptotic behaviour of their 
solutions near the origin $r=0$. By contrast, we are interested in the 
effect of linear perturbations on the subsonic part of the flow ($r>\rso$). 
A single differential equation of second 
order is obtained in Appendix~\ref{Adiff}, with the following structure:
\begin{equation}
{\p^2\over\p r^2}{\delta p\over p}+a_{1}{\p\over\p r}
{\delta p\over  p}
+a_{0}{\delta p\over p}=b_{0} \delta S_R + b_{1} \delta 
K_R.\label{edp}
\end{equation} 
The coefficients $a_{0},a_{1},b_{0},b_{1}$ are functions of $r$ which depend 
on the flow velocity and sound speed radial profiles in the 
unperturbed Bondi flow. 
Their rather lengthy expressions are given in Appendix~B, 
Eqs.~(\ref{coefa} to \ref{coefb}). 
The homogeneous equation corresponding to the left hand side of 
Eq.~(\ref{edp}) describes the propagation of acoustic waves in the 
Bondi flow and their refraction by the sound speed gradients. The 
important feature of Eq.~(\ref{edp}) is the presence of source terms 
on the right hand side: 
\par (i) $\delta S$ is the perturbation of entropy. For the sake of 
simplicity, the ratio of the molecular weight $\mu$ to the gas constant 
${\cal R}$ is set to $\mu/{\cal R} = 1$ throughout this paper, with no loss 
of generality.  
\par(ii) $\delta K$ involves the curl of the vorticity vector $\delta 
w\equiv \nabla\times \delta v $  projected 
along the flow velocity: 
\begin{equation}
\delta K\equiv r^2v\cdot (\nabla\times \delta w)+l(l+1)c^2
{\delta S\over\gamma},\label{defK}
\end{equation}
The radial dependence of the three components of the vorticity vector
$\delta w$ is explicitely integrated in Appendix~\ref{Alinear} 
(Eqs.~\ref{conswr} to \ref{wphi}). Nevertheless, the vorticity 
contributes to the excitation of acoustic waves only through the 
radial part of its curl, as it appears in Eq.~(\ref{defK}).
In an adiabatic flow, both $\delta S$ and $\delta K$ are conserved when 
advected (see Appendix~\ref{Alinear}). A more compact mathematical 
formulation is obtained in Appendix~\ref{Acompact} by writing the 
differential equation satisfied by the perturbation $f$ of the Bernoulli 
constant.
\begin{eqnarray}
\left\lbrace vc^2{\p\over\p r}\left({1-\M^2\over v}
{\p\over\p r}\right)
+{\omega^2-\omega_{l}^{2}\over 1-\M^2}\right\rbrace
\left(\e^{i\omega\int_R^r{\M^2\over1-\M^2}{\d r\over 
v}}f\right)\nonumber\\
={i\omega vc^2\over\gamma}\delta S_R{\p\over\p r}\left\lbrack
{1-\M^2\over \M^2}\e^{i\omega\int_{R}^r{\d r\over 
v(1-\M^2)}}\right\rbrack\nonumber\\
-\delta K_R{ c^2\over r^2}\e^{i\omega\int_{R}^r{\d r\over v(1-\M^2)}},
\label{canonic1}
\end{eqnarray}
where the frequency $\omega_{l}$ is defined as:
\begin{equation}
\omega_{l}^2\equiv l(l+1){c^2-v^2\over r^2}.\label{omegal}
\end{equation}
The structure of this differential equation is the same as 
Eq.~(\ref{edp}), with much simpler coefficients. The right hand side of 
Eq.~(\ref{canonic1}) contains two source terms proportional to 
$\delta S_{R}$ and $\delta K_{R}$. 
The pressure perturbation is related 
to $f$ by the following equation derived in Appendix~\ref{Alinear}:
\begin{equation}
{1\over \rho}{\p \delta p\over \p t} ={{\rm D} f\over {\rm D} 
t}.\label{defp}
\end{equation}

\section{Acoustic waves in the Bondi flow: the refraction cut-off }

\subsection{Acoustic refraction and regularity at the sonic point}

\begin{figure}
\psfig{file=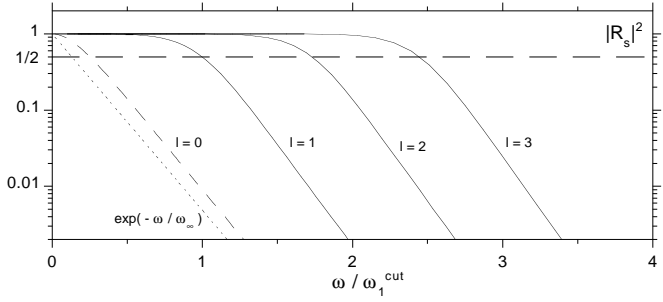,width=\columnwidth}
\caption[]{Refraction coefficient $|\bcmo|^2$ for $\gamma=1.33$, 
$l=0,1,2,3$. The dotted line corresponds to the leading behaviour of 
$|\bcmo|^2$ for $\omega\gg\omcut$ (Eq.~\ref{asympex}) }
\label{figrefrac}
\end{figure}
The homogeneous equation associated with Eq.~(\ref{canonic1}), 
describing the propagation of acoustic waves is:
\begin{eqnarray}
\left\lbrace vc^2{\p\over\p r}\left({1-\M^2\over v}
{\p\over\p r}\right)
+{\omega^2-\omega_{l}^{2}\over 1-\M^2}\right\rbrace
\left(\e^{i\omega\int_R^r{\M^2\over1-\M^2}
{\d r\over v}}f\right)\nonumber\\
=0.\label{homogene}
\end{eqnarray}
This equation is of second order, which reflects the fact that pressure 
perturbations can be decomposed into ingoing and outgoing acoustic waves. 
Such an identification is possible only in the region of the flow where the 
lengthscale of the flow gradients is longer than the wavelength of the 
perturbation, 
where a WKB treatment of the wave propagation can be used (see Appendix~C).
This is always the case far from the accretor in the Bondi flow, 
since the density and sound speed are uniform at infinity. We may 
define $\delta p^{+}$ and $\delta p^-$ as the solutions of the 
homogeneous 
equation propagating respectively inward and outward at infinity.
Following FT2000, the index $+$ (resp. $-$) refers to 
waves propagating in the same (resp. opposite) direction as the flow. 
The acoustic flux carried by these acoustic waves is defined by 
Eq.~(6) of FT2000:
\begin{equation}
F^\pm = {\dot M}_{0}c^{2}{(1\pm\M)^{2}\over\M}
\left|{\delta p^\pm\over\gamma p}\right|^2.\label{aflux}
\end{equation}
Let us normalize $\delta p^\pm_{0}$ so that they carry the same acoustic 
flux at infinity. Any linear combination $\delta p$ of the two independent 
solutions $\delta p^{+}_{0},\delta p^-_{0}$, with complex 
coefficients $\alpha,\beta$, is also a solution of Eq.~(\ref{homogene}):
\begin{equation}
\delta p \equiv \alpha\delta p^-_{0}+\beta\delta p^{+}_{0}.
\label{lincomb}
\end{equation}
There is, however, only one ratio $\alpha/\beta$ such that $\delta p$ 
is regular at the sonic radius. Indeed, a local analysis of the solutions of 
Eq.~(\ref{canonic1}) for $\gamma<5/3$, at the sonic point $\M=1$ using 
Frobenius series, reveals that there is only one solution crossing the 
sonic point regularly, the other exhibiting a logarithmic singularity 
(Bender \& Orszag~1978, Chap.~3.3, p.68). This regularity condition 
reflects the fact that an incoming acoustic wave $\beta\delta p^{+}_{0}$ 
is only partially transmitted to the accretor: an outgoing acoustic wave of 
amplitude $\alpha\delta p^-_{0}$ is refracted out.
Using the same notations as FT2000, the fraction 
$|{\bcmo}|^{2}$ of refracted acoustic flux is deduced from 
Eqs.~(\ref{aflux}) and (\ref{lincomb}):
\begin{eqnarray}
|{\bcmo}|^{2}&\equiv&{F^-\over F^+},\\ 
&=&\left|{\alpha\over\beta}\right|^2.
\end{eqnarray}
The refraction coefficient is therefore directly related to the regularity 
of the solution at the sonic point. 

\subsection{Numerical calculation of the homogeneous solution}

The differential equation (\ref{homogene}) is integrated using a
Runge-Kutta implicit method, from the sonic point to the WKB region. 
The refraction coefficient is then computed by identifying the 
ingoing and outgoing waves in the WKB region.
If $\gamma<5/3$, a Frobenius expansion is used in the vicinity of $\rso$ to 
start the integration away from the singularity.
The calculation for $\gamma=5/3$ and $l=0$ is similar to the case 
$\gamma<5/3$, except that the sonic point is at $r=0$.
If $\gamma=5/3$ and $l\ge1$, the homogeneous solution presents an essential 
singularity at the sonic point, where one branch converges to zero, 
and the other diverges. Nevertheless, an asymptotic expansion of the 
converging solution (see Eqs.~\ref{f530} and \ref{g530})
allows a numerical integration from the vicinity of the origin 
towards infinity. As expected by Kovalenko \& Eremin~(1998), no 
overreflection occurs ($|{\bcmo}|\le 1$). The typical frequency dependence 
of $|\bcmo|^2$ is shown in Fig.~\ref{figrefrac} for $\gamma=1.33$.

\subsection{Cut-off frequency for the refraction of acoustic waves 
for $\gamma<5/3$}

If $\gamma<5/3$, the accretor is surrounded by a supersonic region 
from which acoustic waves cannot escape. Only those waves with a high 
enough frequency may penetrate this region, because lower frequency 
waves are refracted out before the sonic radius. This leads us to expect 
the refraction $|{\bcmo}|^{2}$ to decrease to zero at high frequency, and be
close to unity at low frequency (see Fig.~\ref{figrefrac}). The refraction
cut-off $\omcut$ of a wave of latitudinal order $l$ is defined as the 
intermediate frequency corresponding to a refraction of half of the incoming 
acoustic flux:
\begin{equation}
|\bcmo|^2_{(\omega=\omcut)}\equiv{1\over2}.\label{half}
\end{equation}
Bessel functions are used in Appendix~D to approximate the homogeneous 
solution at high frequency, in order to obtain the leading order of 
the asymptotic behaviour of $|\bcmo|$. This latter decreases exponentially 
above $\omcut$ (see Fig.~\ref{figrefrac}):
\begin{eqnarray}
|\bcmo|(\omega\gg\omcut)&\propto&
\e^{-{\omega\over \omega_\infty}},
\label{asympex}\\
\omega_\infty&\equiv& {\cs|\dMs|\over\pi}={4\over(5-3\gamma)\pi}.
\label{ominf}
\end{eqnarray}

\subsubsection{Non radial acoustic waves $l\ge1$}

\begin{figure}
\psfig{file=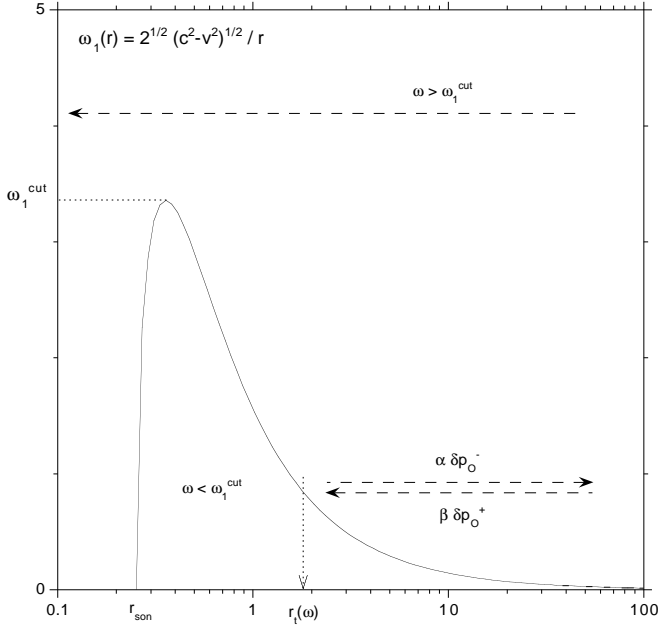,width=\columnwidth}
\caption[]{Typical radial dependence of $\omega_{1}$ (here $\gamma=1.33$), 
showing the turning point $r_{t}$ of a non radial acoustic wave $l=1$ 
below the cut-off frequency $\omega<\omcutu$}
\label{figomcut}
\end{figure}
The turning point of non radial perturbations appears 
explicitly on Eq.~(\ref{homogene}). It corresponds to the radius 
$r_{t}$ solution of $\omega=\omega_{l}(r_{t})$, where $\omega_{l}$ is 
defined by Eq.~(\ref{omegal}) (see Fig.~\ref{figomcut}). For $\gamma<5/3$,
this turning point is double when the frequency equals the 
maximum of the function $\omega_{l}(r_{t})$. The refraction 
coefficient is then equal to $1/2$, according to Bender \& 
Orszag (1978, Chap. 10.6 p. 524):
\begin{eqnarray}
\omcutg&=&{\rm Max}_{\left\{r\ge\rso\right\}}
\left\{\omega_{l}\right\},\label{defcut}\\
&=&\left\lbrack{l(l+1)\over2}\right\rbrack^{1\over2}\omcutu.
\end{eqnarray}
This result was checked by measuring numerically the refraction of 
acoustic waves. Fig.~\ref{figcutoff} shows the dependence of $\omcutg$ on 
the adiabatic index $\gamma$. $\omcutg$ should scale as 
$l^{1/2}(l+1)^{1/2}/ (5-3\gamma)^{5/4}$ according to the asymptotic 
behaviour of $(1-\M^{2})^{1/2}c/r\propto \cs(\dMs/\rso)^{1/2}$ when 
$\gamma\to5/3$, 
deduced from Eqs.~(\ref{radiab} to \ref{cadiab}). A refined comparison with 
the results of numerical calculations of $\omcutg$ leads to
\begin{equation}
\omcutg\sim  l^{1\over2}(l+1)^{1\over2}
\left({2\over 5-3\gamma}\right)^{5\over4},\label{om1}
\end{equation}
over the range $1.4<\cs/c_{\infty}<100$ considered numerically.
The radius $r_{1}$ such that $\omega_{l}(r_{1})=\omcut$ is the minimum 
value of the turning radius $r_{t}$ (see Fig.~\ref{figomcut}). 
$r_{1}(\gamma)$ is independent of $l,m$ and lies just outside the sonic 
surface, and is well fitted by $r_1\sim \sqrt{2}\rso$ for $\gamma$ 
close to $5/3$. 

\subsubsection{Radial acoustic waves $l=0$}

\begin{figure}
\psfig{file=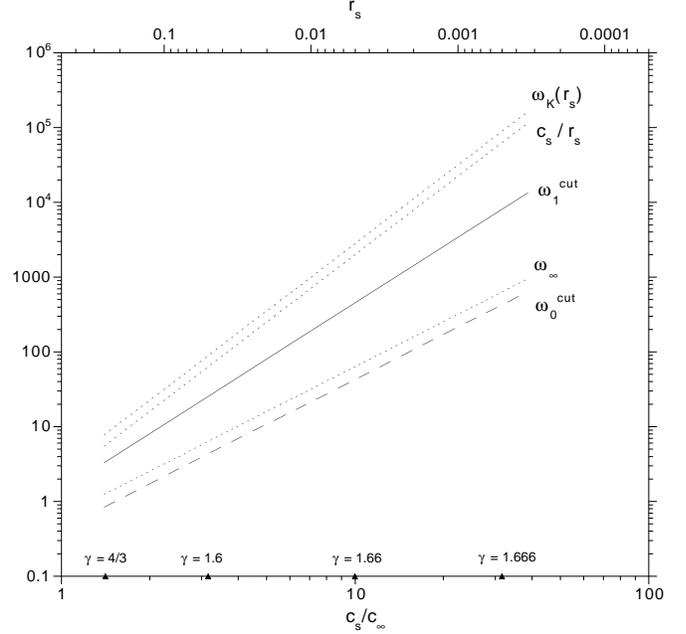,width=\columnwidth}
\caption[]{The cut-off frequencies $\omcutz$ and $\omcutu$ 
correspond to $|\bcmo|^2= 0.5$ for $l=0$ and $l=1$ respectively. They 
cannot be distinguished from their fittings (Eqs.~\ref{om1} and 
\ref{om0}) on this plot. These frequencies are compared to the sonic
frequency $\cs/\rso$, the Keplerian frequency at the sonic radius 
$\omega_{K}$, and $\omega_{\infty}$. Frequencies are expressed in
units of $c_\infty^{3}/GM$, as functions of $\cs/c_{\infty}$ 
(or $\gamma$, or $\rso$) }
\label{figcutoff}
\end{figure}
$\omcutz$ is computed numerically in Fig.~\ref{figcutoff}, and corresponds 
to a lower frequency than 
$\omcutu$. The value of $\omcutz$ can be estimated analytically for 
$\gamma$ close to $5/3$ from the criterion of validity of the WKB 
approximation of acoustic waves, derived in Appendix~C:
\begin{equation}
\omega\gg (1-\M^2)c\left|{\p\log v^2c^2\over\p r}\right|.\label{wkb0}
\end{equation}
Since the acoustic flux is conserved in the region 
of validity of the WKB approximation, any refraction of acoustic waves 
must occur outside this region. We deduce that 
$\omcutz$ scales as the maximum of the function on the right hand 
side of Eq.~(\ref{wkb0}). When $\gamma$ is close to $5/3$, this 
maximum scales as $(1-\M^2)c/r\propto -\dMs\cs\propto (5-3\gamma)^{-1}$.
This is confirmed by numerical calculations of $\omcutz$ which indicate that 
in the range of sound velocities $1.4<\cs/c_{\infty}<100$ studied 
numerically, the ratio $\omcutz/\omega_{\infty}$ is constant within the 
numerical errors:
\begin{equation}
\omcutz\sim {0.85\over5-3\gamma}.\label{om0}
\end{equation}
The scaling of $\omcutz$ should be compared to the two other 
particular frequencies of the flow near the sonic point, $\omcutg$ in 
Eq.~(\ref{om1}) and the local Keplerian frequency $\omega_{K}$:
\begin{equation}
\omega_{K}\sim {8\over(5-3\gamma)^{3\over2}},
\end{equation}
and thus $\omcutz\ll\omcutu\ll\omega_{K}$ for $\gamma$ close to $5/3$. The 
values of $\omcutz,\omcutu$ computed numerically are 
compared to $\cs/\rso$ and to the Keplerian frequency in Fig.~\ref{figcutoff}.

\subsection{Refraction of acoustic waves for $\gamma=5/3$}

In a Bondi flow with $\gamma=5/3$, acoustic waves are perfectly 
refracted at any frequency: 
\begin{equation}
|\bcmo|=1\;{\rm for }\;\gamma={5\over3}.
\end{equation} 
This can be deduced 
from the limit of flows with $\gamma$ close to $5/3$, because the 
cut-off frequency $\omcut$ tends to infinity when $\gamma\to5/3$. 
This is not surprising, since an ingoing acoustic wave of any frequency 
always ultimately meets "a wall", \ie a subsonic region 
where the scale of the flow gradients are much shorter than the 
wavelength.\\
The WKB analysis at high frequency (Appendix~C) indicates that ingoing and 
outgoing acoustic waves are coupled together at $r_0\propto1/\omega$ for 
$l=0$, 
and at their turning point $r_{\rm t}\propto1/\omega^{4\over5}$ if $l\ge1$. 
The continuity between the flow $\gamma=5/3$ and flows with $\gamma$ close 
to $5/3$ is checked by remarking that  $r_{0}\sim \rso$ for 
$\omega\propto (5-3\gamma)^{-1}$ and 
$r_{t}\sim \rso$ for $\omega\propto (5-3\gamma)^{-5/4}$. Comparing this 
to $\omcut$ in Eqs.~(\ref{om0}) and (\ref{om1}), 
the cut-off frequency is simply interpreted as the 
maximum frequency such that the coupling of acoustic perturbations 
occurs in the subsonic flow. 

\section{Acoustic efficiency of entropy and vorticity perturbations}

\subsection{Analytical formulation}

According to Eq.~(\ref{canonic1}), it is natural to define
two dimensionless complex acoustic efficiencies $\cpl$, $\vpl$ 
associated respectively to the two source terms $\delta S_R$ and 
$\delta K_R$. 
These definitions are made so that the acoustic flux $F^-$
of an outgoing sound wave triggered by the perturbations of
entropy and vorticity, measured far from the accretor is:
\begin{equation}
{F^-\over {\dot M}_{0}c_\infty^2}= \left|\cpl\delta S_R + 
{\vpl}{\delta K_R\over c_\infty^2}\right|^2,
\end{equation} 
The integral expressions for $\cpl$ and $\vpl$ are derived in 
Appendix~D:
\begin{eqnarray}
\cpl &=& {i\over 2\gamma\omega}\int_{\rso}^\infty
\e^{\int^r{i\omega\d r\over v(1-\M^2)}}\nonumber\\
&&{\p\over\p r}\left\lbrack{v(1-\M^2)^2\over\M^2}
{\p\over\p r}
\left(f_0\e^{i\omega\int^r{\M^2\over 1-\M^2}{\d r\over v}}\right)
\right\rbrack
\d r,\label{conversionr}\\
\vpl &=& -{i\over2\omega}\int_{\rso}^\infty
{f_0\over r^2 v}
\e^{i\omega\int_R^r {1+\M^2\over 1-\M^2}{\d r\over v}}
\d r.\label{conversionK}
\end{eqnarray}
It is remarkable that both acoustic efficiencies are independent 
of the longitudinal number $m$. $\cpl $ and $\vpl$ depend only on 
$\omega$ and $l$.

\subsection{Numerical calculations}

\begin{figure}
\psfig{file=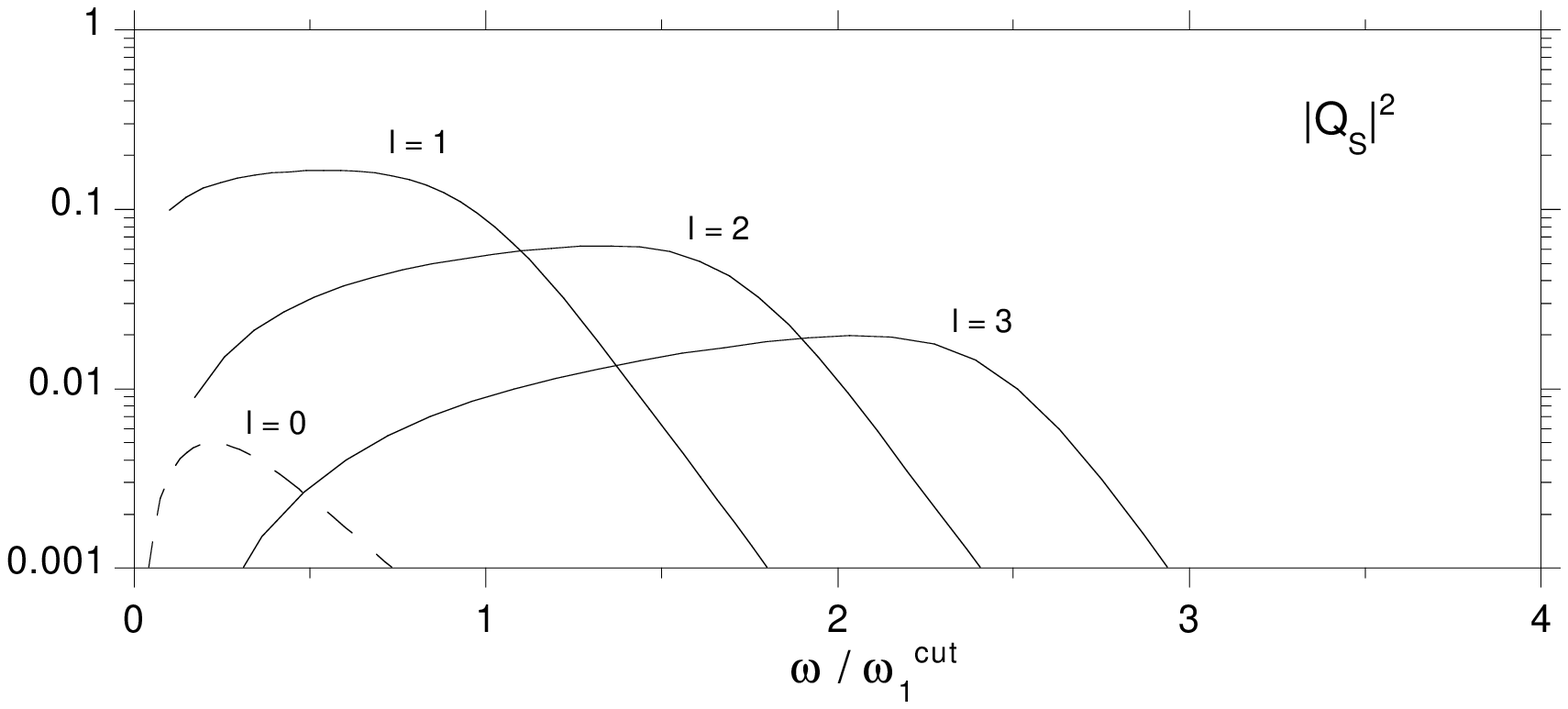,width=\columnwidth}
\psfig{file=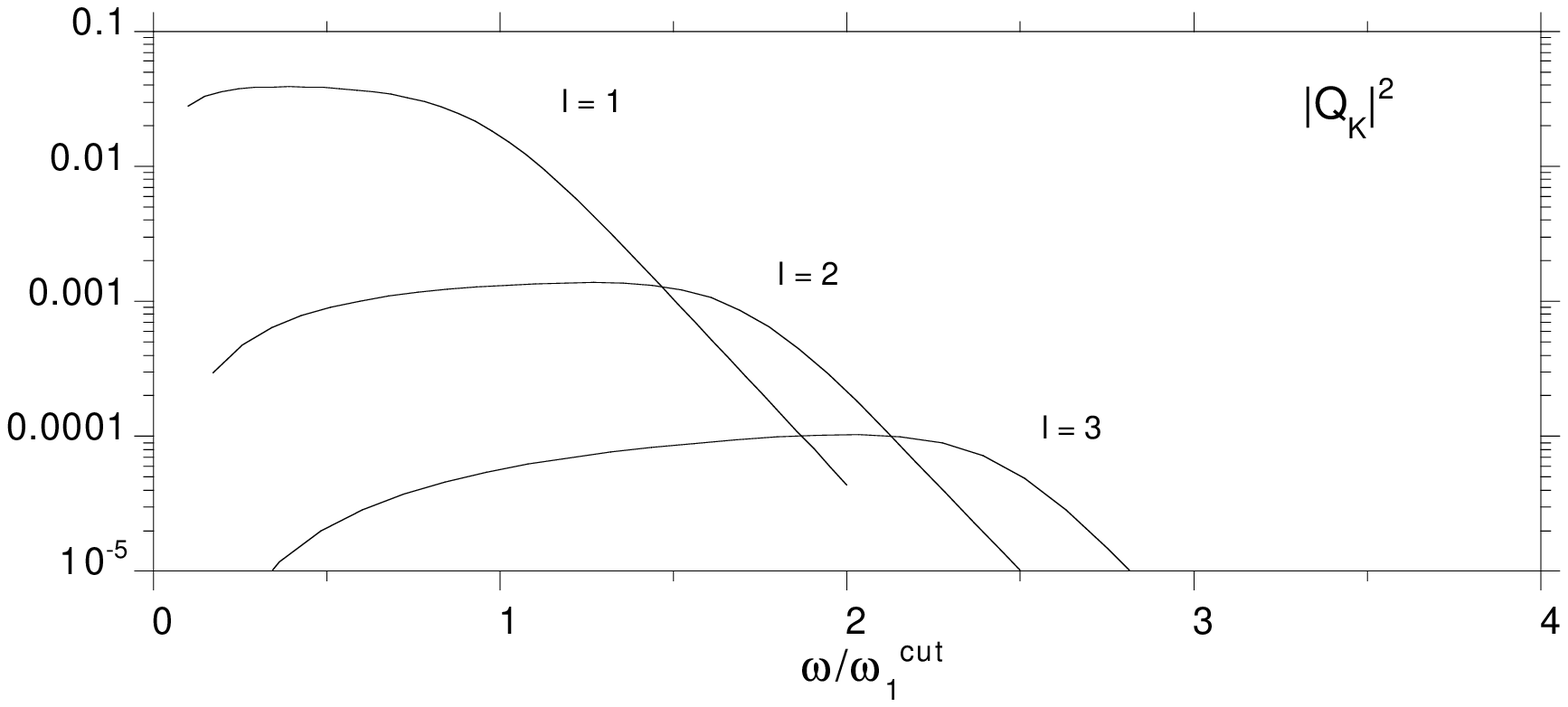,width=\columnwidth}
\caption[]{Acoustic efficiencies $|\cpl|^{2}$ and $|\vpl|^2$ depending 
on the frequency for $\gamma=1.33$, $l=0,1,2,3$}
\label{figcpl1}
\end{figure}
\begin{figure}
\psfig{file=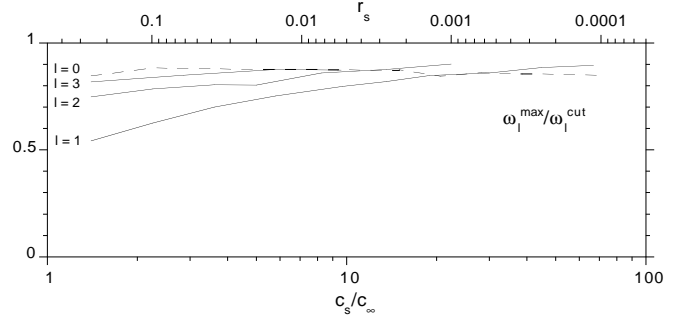,width=\columnwidth}
\caption[]{The frequencies which maximize $|\cpl|$ are
displayed in units of the refraction cut-off $\omcut$, for $l=0,1,2,3$
as functions of $\cs/c_{\infty}$. }
\label{figfreqmax}
\end{figure}
The step of the numerical integration must be able to resolve both 
acoustic and entropy waves. Far from the accretor, the wavelength of 
the acoustic perturbation is asymptotically constant ($\sim 
2\pi c_{\infty}/\omega$), whereas the wavelength of the entropy perturbation 
decreases to zero as $2\pi v/\omega\propto 1/r^{2}$. It is therefore 
useful to transform the integrals in Eqs.~(\ref{conversionr}) and 
(\ref{conversionK}) to accelerate their convergence. This is done 
in Appendix~D by successive integrations by parts, thus obtaining 
Eqs.~(\ref{fastqk}) and (\ref{fastinteg}).
If $\gamma=5/3$, the integrals are transformed near the origin in 
order to correctly treat the singularity.

\subsection{Asymptotic expansion of $|\cpl|$ and $|\vpl|$ at high 
frequency for $\gamma=5/3$ \label{sub53}}

\begin{figure}
\psfig{file=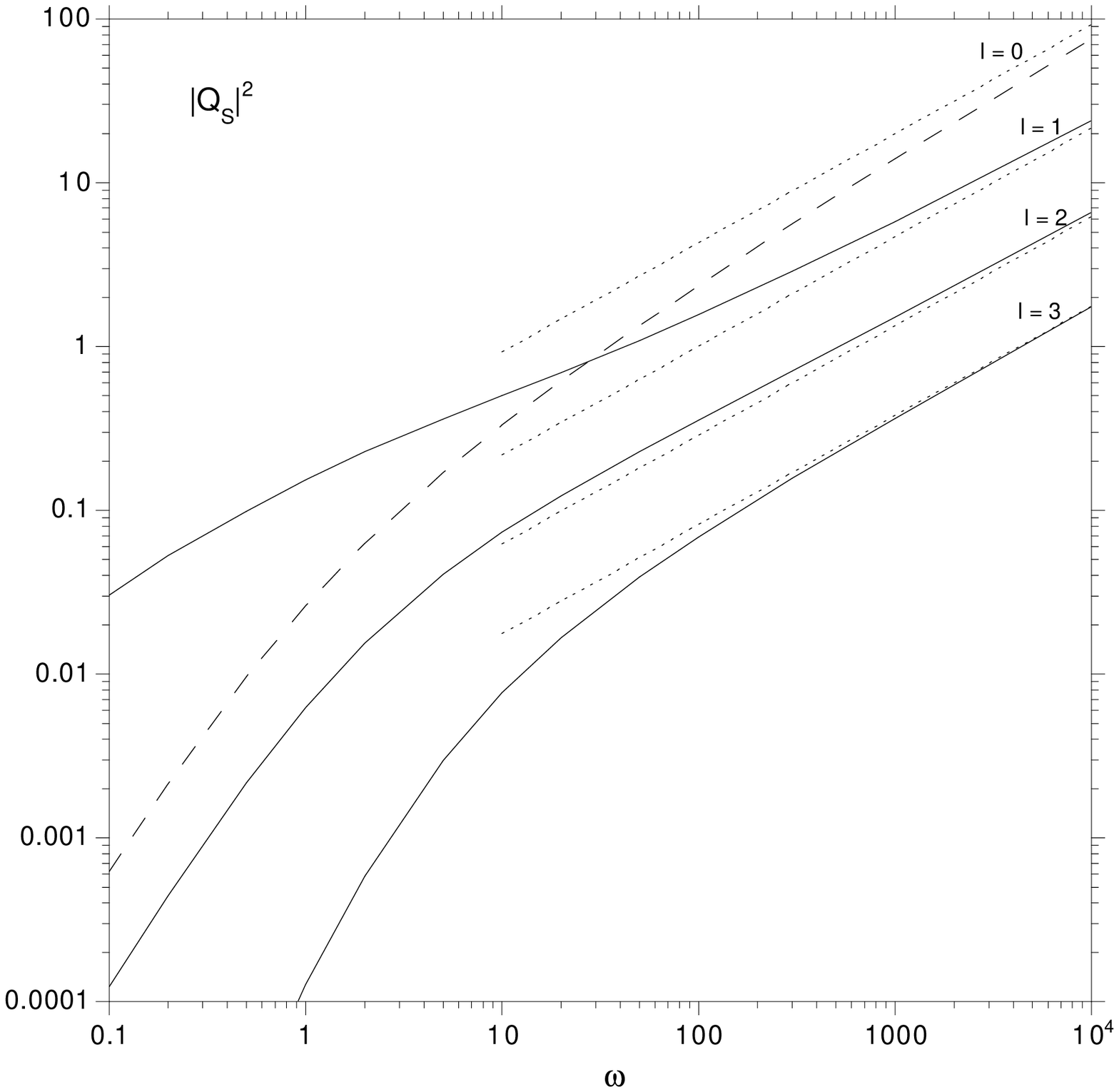,width=\columnwidth}
\caption[]{Acoustic efficiency $|\cpl|^{2}$ of entropy perturbations for 
$\gamma=5/3$, $l=0,1,2,3$. The full line corresponds to the numerical 
calculation, the dotted line corresponds to the asymptotic estimate 
at high frequency (Eqs.~\ref{abscpl} and \ref{nonradial})}
\label{figcpl}
\end{figure}
\begin{figure}
\psfig{file=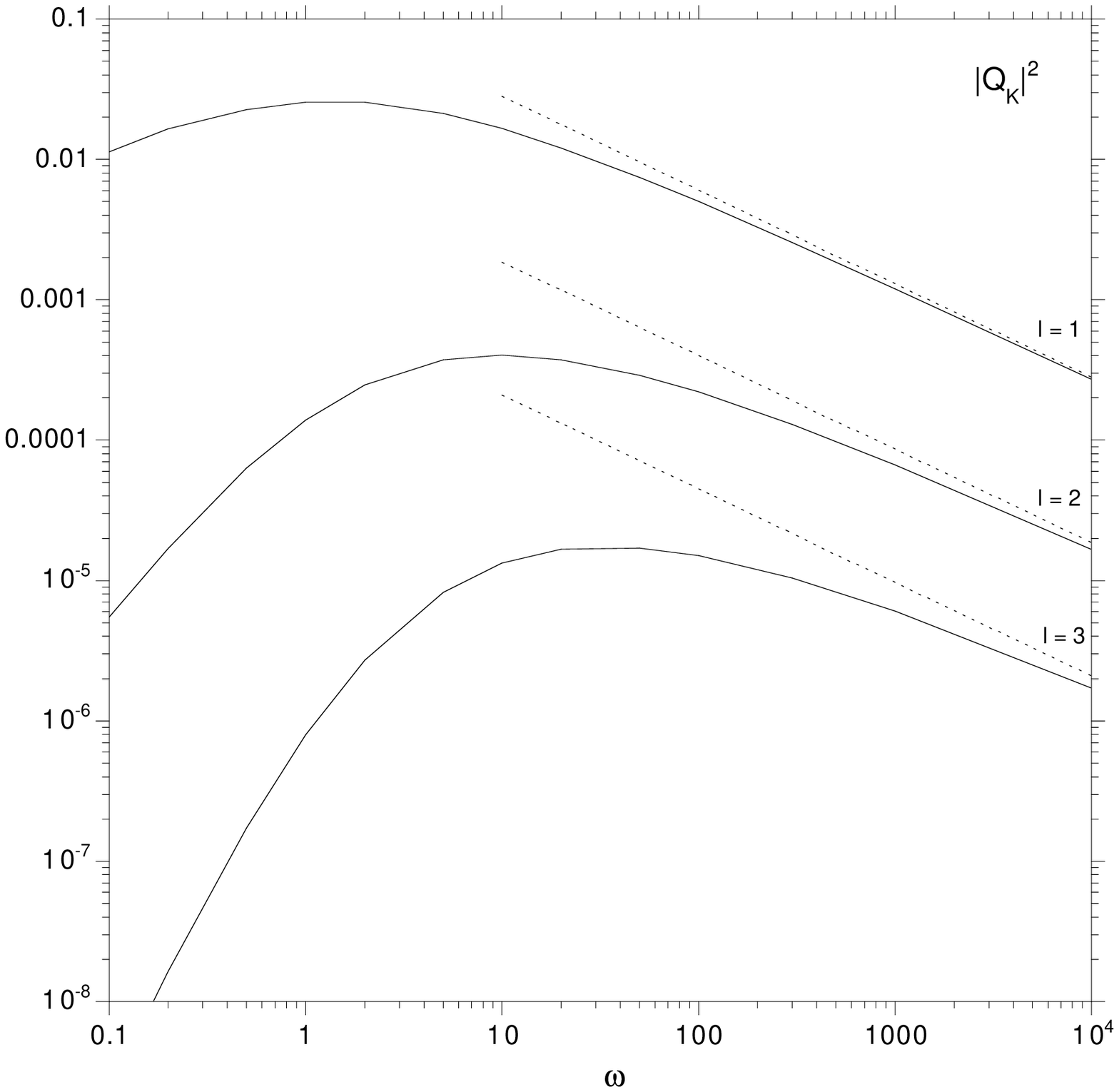,width=\columnwidth}
\caption[]{Acoustic efficiency $|\vpl|^{2}$ of vorticity perturbations for 
$\gamma=5/3$, $l=1,2,3$. The full line corresponds to the numerical 
calculation, the dotted line corresponds to the asymptotic estimate 
at high frequency (Eq.~\ref{asyvpl})}
\label{figvpl}
\end{figure}
Special properties of the Bondi flow $\gamma=5/3$ are related to 
the position of the sonic radius, at the central singularity. 
In particular, both cut-off frequencies
$\omcutz$ and $\omcutu$ are infinite, which means that at any 
frequency, incoming sound waves are totally refracted outward:
\begin{equation}
|\bcmo|=1.
\end{equation}
$|\cpl|$ is estimated in Appendix~\ref{appradial} at high frequency for 
radial perturbations:
\begin{equation}
|\cpl|_{l=0}\sim {3\over5}\left({\omega\over6}\right)^{1\over3}
\Gamma\left({2\over3}\right),
\label{abscpl}
\end{equation}
where the Gamma function satisfies $\Gamma(2/3)\sim 1.354$. \\
The case of non radial perturbations is derived in 
Appendix~\ref{appm}, where we define $L^{2}\equiv l(l+1)$:
\begin{eqnarray}
|\cpl |_{l\ge1} &\sim &{2\over5}
\left({\omega L^2\over2}\right)^{1\over3}
K_{2\over3}\left({2L\over3}\right),
\label{nonradial}\\
|\vpl |_{l\ge1} &\sim&{2\over3}\left({2\over\omega L^2}\right)^{1\over3}
K_{2\over3}\left({2L\over3}\right).\label{asyvpl}
\end{eqnarray}
The asymptotic expansions in Eqs.~(\ref{abscpl}), (\ref{nonradial}) 
and (\ref{asyvpl}), which involve no free scaling parameter, are
successfully compared to numerical calculations of $|\cpl|$  and $|\vpl|$ in 
Figs.~\ref{figcpl} and \ref{figvpl}. 
The radial case of Eq.~(\ref{abscpl}) is recovered by taking the 
continuous limit $l\to 0$ in Eq.~(\ref{nonradial}).
Note that the asymptotic behaviour of the Bessel function 
$K_{2\over3}$
for large arguments ($L\gg1$),
\begin{equation}
K_{2\over3}\left({2L\over3}\right)\sim 
\left({3\pi\over4L}\right)^{1\over2}
\e^{-{2L\over3}},
\end{equation}
makes the acoustic efficiencies decrease rapidly for high order $l$. 
Remembering that the frequency of the entropy perturbation is directly 
related to its radial wavelength ($\lambda\sim 2\pi v/\omega$),
our calculation of $|\cpl|(\omega,l)$ indicates that the 
sound produced by the advection of an object of given mass in the 
Bondi flow depends a lot on its shape. Virtually no sound comes out 
if $l$ is large and $\omega$ is small (\eg a stretched spaghetti).
By contrast, the shape producing the highest acoustic flux corresponds to 
a small $l$ and a large $\omega$ (\eg a wide thin saucer).\\
At high frequency, the coupling of entropy 
and vorticity perturbations to acoustic waves comes essentially from the 
region $r_{\rm eff}\propto 1/\omega^{2/3}$ (\ie the main contribution to the 
integrals in Eqs.~\ref{conversionr} and \ref{conversionK}, computed 
in Appendix~E). The radial scaling at 
high frequency is therefore $r_0\ll r_{\rm t}\ll r_{\rm eff}$.

\subsection{Case of accretion flows with $\gamma$ close to $5/3$}

\begin{figure}
\psfig{file=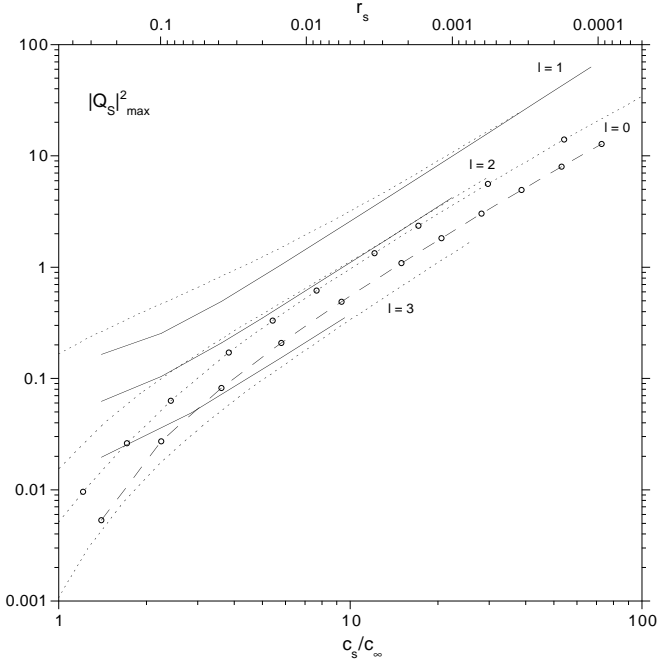,width=\columnwidth}
\caption[]{Maximum acoustic efficiency $|\cpl|^{2}$ as a 
function of $\cs/c_{\infty}$ (\ie of $\gamma<5/3$) for $l=0,1,2,3$. The 
dotted lines are deduced from the curves $|\cpl|^{2}(\omega)$, $\gamma=5/3$
(Fig.~\ref{figcpl}), where $\omega$ is replaced by 
$\omega_{l}^{\rm max}\sim 0.8\omcut$. Curves corresponding to $l=0$ 
are marked with circles}
\label{figcpl2}
\end{figure}
\begin{figure}
\psfig{file=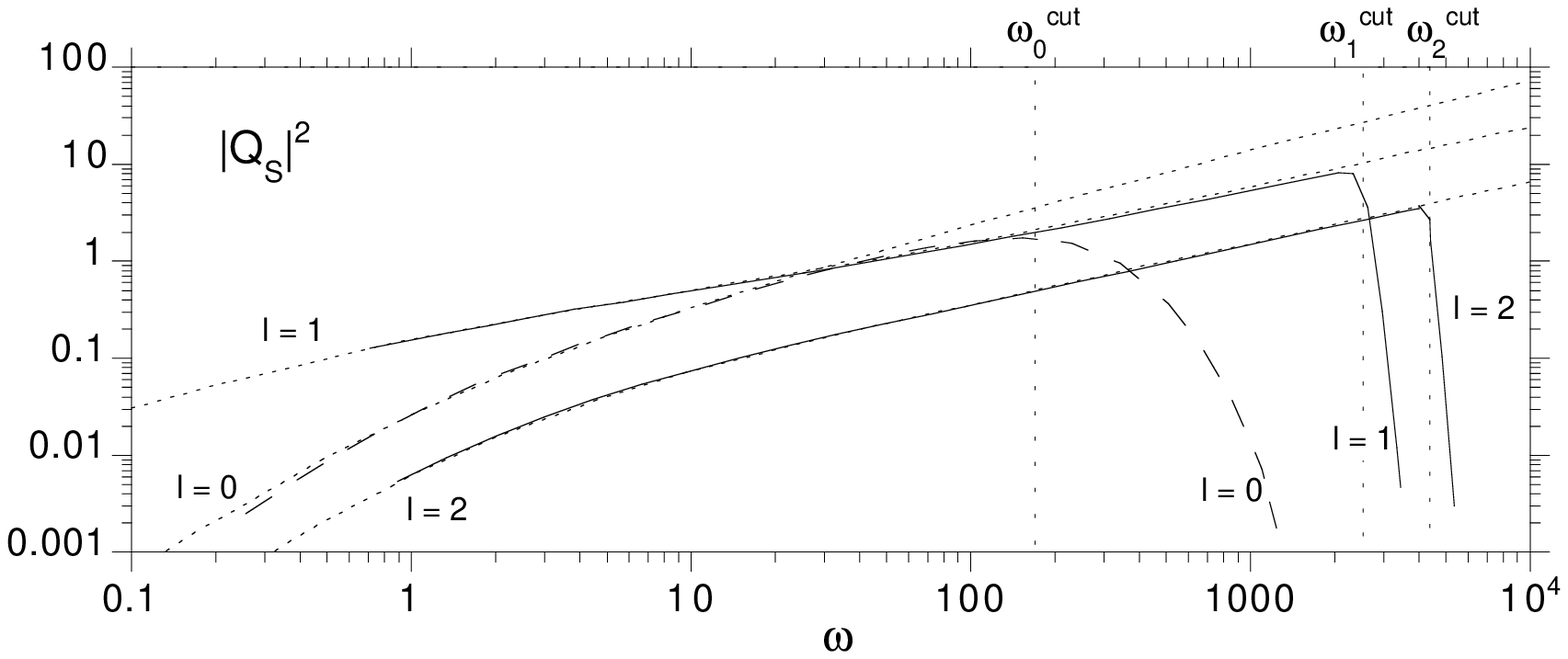,width=\columnwidth}
\psfig{file=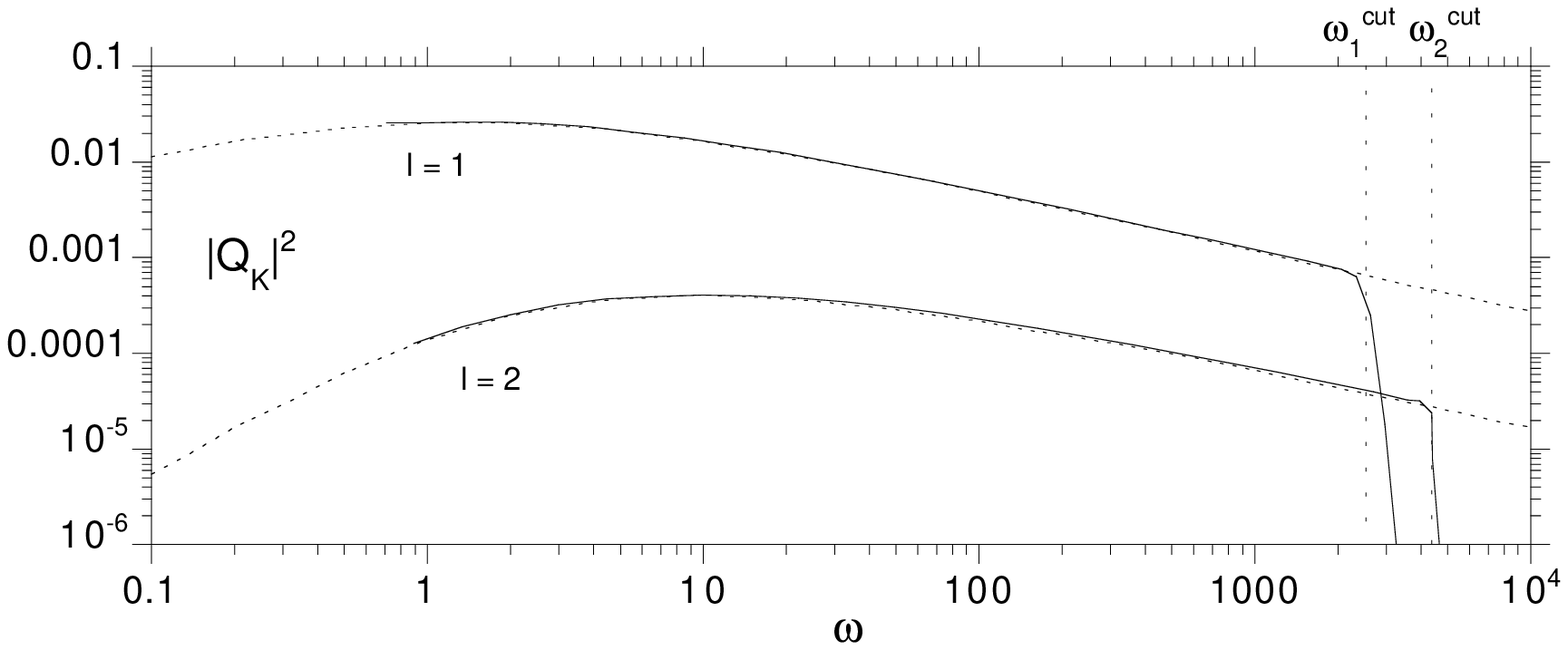,width=\columnwidth}
\caption[]{Acoustic efficiencies $|\cpl|^{2}$ and $|\vpl|^2$ depending on the 
frequency for $\gamma=1.665$ (\ie $\cs/c_{\infty}=20$), $l=0,1,2$.
Frequencies are expressed in units of $c_{\infty}^{3}/GM$. 
These efficiencies are very close to those obtained for 
$\gamma=5/3$ (dotted lines), truncated after the cut-off frequency 
$\omcut$.}
\label{figcpl20}
\end{figure}
For $\gamma$ close to $5/3$ and at frequencies lower than the cut off 
frequency, we can use the approximation of $|\cpl|$ obtained for 
$\gamma=5/3$. 
According to Fig.~\ref{figfreqmax}, the maximum of $|\cpl |$ is reached 
at the frequency $\omega_{l}^{\rm max}$, with
$0.5<\omega_{l}^{\rm max}/\omcut\le 1$. The asymptotic behaviour of 
the maximum value of $|\cpl|$ 
is therefore estimated by extrapolating Eqs.~(\ref{abscpl}) and 
(\ref{nonradial}) to the cut-off frequency:
\begin{eqnarray}
{\rm Max}|\cpl|_{l=0}&\sim  &{1\over\gamma}
\left({\omcutz\over6}\right)^{1\over3}\Gamma\left({2\over3}\right),\\
&\propto& \rso^{-{1\over3}},\\
{\rm Max}|\cpl|_{l\ge1}&\sim & 
{2^{1\over2}\over3\gamma} (\omcutu)^{1\over3}
LK_{2\over3}\left({2 L\over3}\right),\\
&\propto& L K_{2\over3}\left({2L\over3}\right)
\rso^{-{5\over12}}.\label{cplasympt}
\end{eqnarray}
It is interesting to note in Fig.~\ref{figcpl2} that the most efficient 
entropy perturbations at a given frequency when $\gamma<5/3$ are non radial 
($l=1$), whereas radial ones are more efficient for $\gamma=5/3$ 
(Fig.~\ref{figcpl}). This is directly related to 
the fact that the critical refraction frequency is much higher for $l=1$ 
than for $l=0$ (Eqs.~\ref{om1} and \ref{om0}). This is illustrated in 
Fig.~\ref{figcpl20}, where the acoustic efficiencies computed for 
$\cs/c_{\infty}=20$ ($\gamma=1.665$) are very close to 
their values for $\gamma=5/3$, up to the refraction cut-off. In this 
figure, the efficiency $|\cpl|$ is higher for $l=0$ than for $l=1$ in the 
limited range of frequencies $30<\omega<120 $, but the overall maximum is 
reached near $\omcutu\sim 2300$ for $l=1$. 

\section{Conclusions \label{Sconclusion}}

\subsection{Summary}

The acoustic response of the Bondi flow to pressure, entropy and 
vorticity perturbations has been studied thoroughly.
Let us summarize the main results:
\par (i) the critical refraction frequency for acoustic waves 
$\omcut$ is quite different from the "natural" frequency 
$\cs/\rso$ one might think of at first glance. Moreover $\omcut$ is also very 
different for radial and non radial perturbations. 
The two fundamental frequencies $\omcutz,\omcutu$ of this flow near the 
sonic point were accurately determined (Eqs.~\ref{om1} and \ref{om0}), and
are ordered as follows: $\omcutz<\omcutu<\omega_{K}$, with
\begin{eqnarray}
\omcutz&\propto& {\rm Max }\left\lbrace(1-\M^{2}){c\over r}\right\rbrace
\propto \left({\p\M\over\p\log r}\right)_{\rm s}{\cs\over\rso},\\
\omcutu&\propto& {\rm Max }\left\lbrace(1-\M^{2})^{1\over2}
{c\over r}\right\rbrace
\propto \left({\p\M\over\p\log r}\right)_{\rm s}^{1\over2}{\cs\over\rso}.
\end{eqnarray}
\par (ii) two quantities conserved in the linear approximation 
of the perturbed Bondi flow are source terms for acoustic waves: the 
entropy perturbation $\delta S$, and
a quantity $\delta K$ related to both vorticity and entropy 
perturbations defined by Eq.~(\ref{defK}).
\par (iii) the acoustic efficiencies $\cpl$ and $\vpl$ of 
entropy and vorticity perturbations are independent of 
the longitudinal number $m$.
\par (iv) $|\cpl|$ and $|\vpl|$ are highest for $l=1$ perturbations
if $\gamma<5/3$.
\par (v) $|\vpl|$ is highest at relatively low frequency, and is 
always rather moderate ($|\vpl|<1$).
\par (vi) $|\cpl|$ is highest at high frequency near the 
refraction cut-off, and can be much larger than one if $\gamma$ is 
close to $5/3$, \ie if $\cs/c_{\infty}\gg1$. This confirms the physical 
argument of FT2000 which stressed the importance of the enthalpy increase 
between the outer part of the flow and the sonic point.
\par (vii) asymptotic scalings of $|\cpl|,|\vpl|$ were obtained at 
high frequency for $\gamma=5/3$ (Eqs.~\ref{abscpl} to \ref{asyvpl}), 
in excellent agreement with numerical calculations.\\
The analytical expressions obtained for $\omcut(\gamma)$, and for
$|\cpl|$ and $|\vpl|$ at high frequency for $\gamma=5/3$, 
are accurate enough (see Figs.~\ref{figcutoff}, \ref{figcpl} 
and \ref{figvpl}) to be used as tests of the accuracy of 3D hydrodynamical 
codes used for Bondi accretion, and particularly their limitations at 
high frequency.\\

These results suggest that if a shock were present as outer 
boundary, the entropic-acoustic cycle would be unstable with respect 
to high frequency waves if the sound speed at the sonic point is high 
enough,\ie if $\gamma$ is close to $5/3$. Indeed, an entropy 
perturbation $\delta S_{1}$ advected towards the accretor triggers an 
outgoing acoustic flux $F^{-}={\dot M}_{0}c^{2}|\cpl|^{2}|\delta S|^{2}$. 
These acoustic waves propagate against the flow until they reach the shock,
where they produce new entropy perturbations $\delta S_{2}$ 
(see FT2000), with $\delta S_{2}>\delta S_{1}$ since $|\cpl|\gg1$. 
The eigenmodes corresponding to this unstable cycle will be investigated 
in the second paper of this series (Foglizzo 2001). The fact that the 
most unstable simulations of Bondi-Hoyle-Lyttleton accretion 
also correspond to $\gamma=5/3$ (Ruffert \& Arnett 1994, Ruffert 1994b) 
already encourages us to look for an extrapolation of these results to the 
case of non-radial shocked accretion flows.

\subsection{Discussion}

\subsubsection{Upper bound for $|\cpl|^{2}$ imposed by the Schwarzschild 
radius}

In the flow $\gamma=5/3$, the fact that the outgoing acoustic flux 
produced by entropy perturbations diverges at high frequency 
(Eqs.~\ref{abscpl} and \ref{nonradial}) is rather 
surprising, and directly related to the hypothesis of a point-like 
newtonian accretor ($\rso=0$).
In this section we show that $|\cpl|^{2}$ is naturally bounded
if the finite size of the accretor is taken into account. 
Simple arguments are used in order to roughly estimate this upper bound in 
the case of an accreting black hole, without computing the exact relativistic 
corrections near its horizon.\\
Let us denote by $r_{\rm Sch}=2GM/\cl^{2}$ the Schwarzschild radius 
of the black hole, where $\cl$ is the speed of light.
With a sound speed at infinity $c_{\infty}\ll\cl$, Eq.~(\ref{rson}) 
guarantees that relativistic effects are negligible except for $\gamma$ 
close to $5/3$. Indeed Petrich \etal (1989) checked numerically 
that relativistic corrections are small for $\gamma=1.1$ and $\gamma=4/3$.
If $\gamma=5/3$, the sonic point is bound to lie 
at a few Schwarzschild radii:
\begin{equation}
\eta\equiv {\rso\over r_{\rm Sch}}>1.
\end{equation}
Our description of non radial acoustic waves in the flow $\gamma=5/3$ is not 
affected by relativistic corrections as long as their turning point $r_{t}$ 
is far enough from the horizon. Since $r_{t}\sim (L/\omega)^{4/5}$ at high 
frequency (Eq.~\ref{turn}), a natural relativistic cut-off is 
introduced for $r_{t}\sim \eta r_{\rm Sch}$:
\begin{equation}
\omcutg\sim 
{0.25\over\eta^{5\over4}}\left\lbrack{l(l+1)\over2}\right\rbrack^{1\over2}
\left({c_{\infty}\over \cl}\right)^{1\over2}{c^{3}\over GM}.
\end{equation}
The efficiency $|\cpl|^{2}$ of the generation of non radial acoustic
waves close to this cut-off frequency is deduced from 
Eq.~(\ref{nonradial}):
\begin{equation}
|\cpl|^{2}\sim {0.1\over\eta^{5\over6}}\left({L^{2}\over2}\right)^{7\over6}
\e^{-{4L\over3}}\left({\cl\over c_{\infty}}\right)^{5\over3}.
\label{expo}
\end{equation}
The numerical values of the cut-off frequency and maximum acoustic 
efficiency of the entropy perturbations $l=1$ are thus typically:
\begin{eqnarray}
\omcutu&\sim&{100\over\eta^{5\over4}} 
\left({c_{\infty}\over 100\;{\rm km/s}}\right)^{1\over2}
\left({M\over 10 M_{\rm sol}}\right)^{-1}\;{\rm Hz} ,\\
|\cpl|^{2}_{\rm max}&\sim&{10^{4}\over \eta^{5\over6}}
\left({c_{\infty}\over 100\;{\rm 
km/s}}\right)^{-{5\over3}}.\label{numqs}
\end{eqnarray} 
Therefore relativistic effects set an upper bound on the apparent 
divergence of $|\cpl|$ at high frequency for $\gamma=5/3$. 
Eq.~(\ref{numqs}) indicates that entropy perturbations may excite 
low degree acoustic waves very efficiently, up to non linear amplitudes
if the component $l=1$ of the entropy perturbations exceeds a few percent.
This efficiency is much smaller for inhomogeneities with a small 
angular scale, since $|\cpl|$ decreases exponentially when $l$ increases 
(Eq.~\ref{expo}). 

\subsubsection{Non adiabatic limitations}

Non adiabatic processes are discussed by Chang \& Ostriker (1985) in the 
context of Bondi accretion. The efficiency $|\cpl|$ would be strongly 
affected by a strong thermal conduction which smoothes out
entropy perturbations. This effect would be more pronounced for short 
wavelength perturbations, which are also the most efficient for $\gamma$ 
close to $5/3$. Without 
going into the details of such processes, our understanding of the 
acoustic efficiency allows the following remark: if the inner 
region of the flow were dominated by non adiabatic processes, the optimal 
frequency for $|\cpl|$ would be at most reduced to the maximum frequency of 
acoustic waves refracted inside the adiabatic region of the flow. 
Even when there is no shock, the mechanism of excitation of acoustic waves 
from the advection of entropy perturbations opens interesting perspectives 
to the problem of time delay in Cygnus X-1, approached numerically by 
Manmoto \etal (1996). Interpreting their results in terms of acoustic 
efficiency of entropy perturbations suggest that this physical process 
survives the inclusion of rotation, radiative cooling and viscous 
heating. The study of effects deserve a careful analysis which is beyond 
the scope of the present paper.

\acknowledgements
The author acknowledges stimulating discussions with Michel Tagger.
The Runge-Kutta algorithm used in this work for the numerical calculations
was kindly provided by Roland Lehoucq.

\appendix
\section{Description of the unperturbed Bondi flow\label{unperturbed}}
\subsection{$\gamma<5/3$}

The Bernoulli equation and the conservation of mass are written after 
normalizing velocities to the sound speed $c_\infty$ at infinity, and 
distances to the Bondi radius $GM/c_\infty^2$
\begin{eqnarray}
{v^2\over2}+{c^2-1\over\gamma-1} &=& {1\over r},\label{bernou}\\
-r^2vc^{2\over\gamma-1}&=& 
\left({1\over2}\right)^{\gamma+1\over2(\gamma-1)}
\left({4\over5-3\gamma}\right)^{5-3\gamma\over2(\gamma-1)}.\label{consmass}
\end{eqnarray}
By taking the radial derivative of these equation, we obtain:
\begin{eqnarray}
{\p\log c\over\p\log r}&=&{\gamma-1\over2}{\M^2\over1-\M^2}
\left(2-{1\over rv^2}\right),\label{derivc}\\
{\p\log v\over\p\log r}&=&{1\over1-\M^2}
\left({1\over rc^2}-2\right).\label{derivv}
\end{eqnarray}
At the sonic point, $\rso,\cs$ are defined by Eqs.~(\ref{rson}) and 
\ref{cson}). A Taylor expansion of Eqs.~(\ref{derivc}) and 
(\ref{derivv}) leads to:
\begin{eqnarray}
{\p\log v\over\p\log r}(\rso)&=&-{2\over\gamma+1}\left\lbrack
\gamma-1+\left({5-3\gamma\over2}\right)^{1\over2}\right\rbrack,\\
{\p\log c\over\p\log r}(\rso)&=&-{\gamma-1\over\gamma+1}\left\lbrack
2-\left({5-3\gamma\over2}\right)^{1\over2}\right\rbrack,\\
{\p\log \M\over\p\log r}(\rso)&=&-\left({5-3\gamma\over2}\right)^{1\over2}.
\end{eqnarray}

\subsection{$\gamma=5/3$\label{Aadiab}}

In accretion flows where $\gamma=5/3$, Eqs.~(\ref{bernou}) and 
(\ref{consmass}) become:
\begin{eqnarray}
{v^2\over2}+{3\over2}(c^2-1)&=&{1\over r}, \\
r^2vc^3 &=& -{1\over 4}.
\end{eqnarray}
$r,v,c$ can be expressed as explicit functions of the variable $x$ defined by:
\begin{eqnarray}
x&\equiv &{1\over\M^{1\over2}},\\
{\p r\over\p x}&=&{1-\M^2\over2},\\
r&=&{(x-1)^2(9x^4+2x^2+1)\over6x^3(3x^2-2x+1)},\label{radiab}\\
v^2&=&{3\over(x-1)^2}{3x^2-2x+1\over9x^4+2x^2+1},\\
c^2&=&{3x^4\over(x-1)^2}{3x^2-2x+1\over9x^4+2x^2+1}.\label{cadiab}
\end{eqnarray}
 
\section{Linearized equations for perturbations}

\subsection{Second order differential system\label{Alinear}}

The Euler equation is written as follows:
\begin{equation}
{\p v\over\p t}+w\times v + \nabla\left({v^2\over2}+{c^2\over\gamma-1}
-{GM\over r}\right)=c^2\nabla{S\over\gamma}.\label{Euler}
\end{equation}
Projecting the Euler equation onto the flow velocity, we obtain the 
equation of evolution of the Bernoulli constant:
\begin{equation}
\left({\p\over\p t}+v\cdot\nabla\right)
\left({v^2\over2}+{c^2\over\gamma-1}
-{GM\over r}\right) = {1\over\rho}{\p p\over\p t}.\label{dber}
\end{equation}
By combining the curl of Euler equation and the mass conservation, we 
obtain the equation for the evolution of the vorticity $w$:
\begin{equation}
{\p\over\p t}{w\over\rho}+({v\cdot\nabla}){w\over\rho}=
\left({w\over\rho}\cdot\nabla\right)v+{1\over\rho}\nabla 
c^2\times\nabla {S\over\gamma}.
\label{vort}
\end{equation}
We make a Fourier transform in time of the linearized equations 
describing the radial flow perturbed in spherical coordinated 
$(r,\theta,\varphi)$.
The linearized equation of entropy conservation can be directly 
integrated:
\begin{equation}
\delta S= \delta S_R\eiwv.
\end{equation}
The vorticity equation (\ref{vort}) can also be integrated when 
linearized:
\begin{eqnarray}
w_r&=&\left({R\over r}\right)^2 (w_r)_R\eiwv,\label{conswr}\\
w_\theta&=&{1\over rv}\left\lbrack Rv_R(w_\theta)_R -
{c^2-c_R^2\over\sin\theta}{\p\over\p\varphi}{\delta 
S_{R}\over\gamma}\right\rbrack\eiwv,\label{wtet}\\
w_\varphi&=&{1\over rv}\left\lbrack Rv_R(w_\varphi)_R +
(c^2-c_R^2){\p\over\p\theta}
{\delta S_{R}\over\gamma}\right\rbrack\eiwv.\label{wphi}
\end{eqnarray}
Thus the product $r^{2}w_{r}$ is conserved during advection, as 
remarked by Kovalenko \& Eremin (1998).
In order to write the linearized Euler equation in the simplest form, 
let us define the two functions $f,g$ as follows:
\begin{eqnarray}
f&\equiv& v\;\delta v_r + {2\over\gamma-1} c\;\delta c,\label{deff}\\
g&\equiv& {\delta v_r\over v} + {2\over\gamma-1}{\delta c\over 
c}.\label{defg}
\end{eqnarray}
$f$ is the perturbation of the Bernoulli constant, which is directly 
related to the pressure variations according to Eq.~(\ref{dber}). $g$ 
is related to the perturbation of the mass accretion rate and entropy 
through:
\begin{equation}
g={\delta \dot M\over \dot M} + \delta S.
\end{equation}
By linearization of Eq.~(\ref{Euler}), we express the velocity 
components
$(\delta v_\theta, \delta v_\varphi)$ as follows:
\begin{eqnarray}
{\delta v_\theta\over v}&=&{w_\varphi\over i\omega}+
{1\over i\omega rv}
{\p\over\p\theta}f-{c^2\over i\omega rv}{\p\over\p\theta}{\delta 
S_{R}\over\gamma}\eiwv,\\
{\delta v_\varphi\over v}&=&-{w_\theta\over i\omega}+
{1\over i\omega rv\sin\theta}\left\lbrack
{\p\over\p\varphi}f-c^2{\p\over\p\varphi}{\delta S_{R}\over\gamma}
\eiwv\right\rbrack.
\end{eqnarray}
The radial part of the Euler equation, together with the continuity 
equation, lead to the following differential system:
\begin{eqnarray}
v{\p f\over\p r}+{i\omega \M^2f\over 1-\M^2}
&=& {i\omega v^2 g\over 1-\M^2} + i\omega c^2 {\delta 
S_R\over\gamma}\eiwv,\\
v{\p g\over\p r}+{i\omega \M^2g\over 1-\M^2}
&=& {i\omega f\over c^2(1-\M^2)} + 
{i\over \omega}\Delta_{\theta,\varphi}f+{i\delta K_R\over r^2\omega 
}\eiwv,\nonumber\\
&&
\end{eqnarray}
where $\Delta_{\theta,\varphi}$ is the non radial part of the 
Laplacian in spherical coordinates. The constant $\delta K_R$ is 
defined as follows:
\begin{equation}
\delta K\equiv r^2\left\lbrack v\cdot (\nabla\times w)
-c^2\Delta_{\theta,\varphi}{\delta S\over\gamma}\right\rbrack.
\end{equation}
We deduce from Eqs.~(\ref{wtet}-\ref{wphi}) that $\delta K$ is conserved 
when advected:
\begin{equation}
\left\lbrace{\p\over\p t}+v{\p\over\p r}\right\rbrace\delta K=0.
\end{equation}
$\delta K$ can be written independently of the system of coordinates, 
as the first order perturbation of the quantity $K$ defined by:
\begin{equation}
K\equiv {{\dot M}\over4\pi}{{\bf v}\over\rho v}.\left\lbrack \nabla\times 
{\bf w} - c^{2}
\nabla\times\left( {{\bf v}\over v^{2}}\times\nabla {S\over\gamma}\right)
\right\rbrack,
\end{equation}
but the conservation of $K$ is established only in the linear approximation.
By projecting $f,g,\delta S_R$ and the radial component of the curl 
of the vorticity $(\nabla\times w_R)_r$ onto the spherical 
harmonics $Y_l^m(\theta,\varphi)$, which are the eigenvectors of the 
Laplacian, we obtain:
\begin{eqnarray}
v{\p f\over\p r}+{i\omega \M^2f\over 1-\M^2}
&=& {i\omega v^2 g\over 1-\M^2} + i\omega c^2 {\delta 
S_R\over\gamma}\eiwv,
\label{dfdr}\\
v{\p g\over\p r}+{i\omega \M^2g\over 1-\M^2}
&=& {i\omega f\over c^2(1-\M^2)} - {iL^{2}\over \omega r^2}f+
{i\delta K_R\over r^2\omega }\eiwv,\label{dgdr}
\end{eqnarray}
where we have defined $L^2\equiv l(l+1)$ for the sake of the simplicity 
of the equations.

\subsection{Second order differential equation\label{Adiff}}

\subsubsection{Pressure perturbation\label{Apress}}

The most natural function to describe the behaviour of acoustic waves 
is the pressure perturbation $\delta p/p$, which is related to the sound 
speed perturbation through:
\begin{equation}
{\delta p\over\gamma p}={2\over\gamma-1}{\delta c\over c} - 
{\delta S_R\over\gamma}\eiwv.
\end{equation}
Using this equation in Eqs.~(\ref{deff}-\ref{defg}) enables us to 
write $(f,g)$ as functions of $\delta v_r$ and $\delta p$ in 
Eqs.~(\ref{dfdr}-\ref{dgdr}), and thus obtain Eq.~(\ref{edp}) with:
\begin{eqnarray}
a_{1}&\equiv&{\p\log\over\p r}{(c^2-v^2)^2\over v\Delta } + 
{2i\omega v\over c^2-v^2},\label{coefa}\\
a_{0}&\equiv&{1\over c^2-v^2}\left\lbrace\omega^2-L^{2}{c^2\over r^2}
-{1\over v}{\p\over\p r}\left({c^4\over v}{\p\M^2\over\p 
r}\right)+\right.\nonumber\\
&&\left.2i\omega{\p v\over\p r}
+c^2{\p\log\Delta\over\p r}\left\lbrack
{\p\log\M^2\over\p r}-(1+\M^2){i\omega\over 
v}\right\rbrack\right\rbrace,\\
b_{0}&\equiv& {-i\Delta\over vc^2(1-\M^2)}\eiwv
{\p\over\p r}\left\lbrack
{c^{2}\over\Delta}\left(\omega -iv{\p\log\over\p r}{1\over 
\M^2}\right)\right\rbrack,\\
b_{1}&\equiv& -{i\gamma\Delta\over vc^2(1-\M^2)}\eiwv
{\p\over\p r}{v^2\over\omega r^2\Delta},\label{coefb}\\
\Delta&\equiv& \omega^2+ 2i\omega{\p v\over\p r} + L^{2}{v^2\over r^2}.
\end{eqnarray}
The function $\Delta$ is introduced by our choice of writing the 
differential equation satisfied by the pressure perturbation, which 
is justified from the physical point of view. Note that $\Delta$ 
never vanishes as long as the real part of the frequency is different 
from zero. 

\subsubsection{A more compact mathematical formulation
\label{Acompact}}

It is convenient to define ($\tilde f,\tilde g$) as:
\begin{eqnarray}
{\tilde f}&\equiv&\e^{i\omega\int_R^r{\M^2\over1-\M^2}{\d r\over 
v}}f,\label{ftilde}\\
{\tilde g}&\equiv&\e^{i\omega\int_R^r{\M^2\over1-\M^2}{\d r\over v}}g.
\end{eqnarray}
The differential system Eqs.~(\ref{dfdr}-\ref{dgdr}) is then simpler:
\begin{eqnarray}
{\p \tilde f\over\p r}&=& {i\omega v {\tilde g}\over1-\M^2} 
+ i\omega{c^2\over v}
{\delta S_R\over\gamma}\e^{i\omega\int_R^r {\d r\over v(1-\M^2)}},
\label{dtfdr}\\
{\p \tilde g\over\p r}&=& {i{\tilde f}\over \omega v}
\left\lbrack{\omega^2\over c^2(1-\M^2)} - {L^{2}\over 
r^2}\right\rbrack
+{i\delta K_R\over r^2\omega v}
\e^{i\omega\int_R^r {\d r\over v(1-\M^2)}}.\label{dtgdr}
\end{eqnarray}
The homogeneous differential equation~( \ref{homogene}) can
be transformed into a more compact form using the new variable $X$:
\begin{eqnarray}
{\d X\over\d r}&\equiv& {v\over 1-\M^2},\\
W &\equiv& {1\over v^2c^{2}}(\omega^2-\omega_{l}^{2}),\label{defW}
\end{eqnarray}
where $\omega_{l}$ is defined by Eq.~(\ref{omegal}).
\begin{eqnarray}
{\p^2 \tilde f\over\p X^2}+W \tilde f=\nonumber\\
-{1-\M^2\over v}\e^{i\omega\int{\d X\over v^2}}\left\lbrace
{\omega\over \M^2}{\delta S_R\over \gamma}\left( {\omega\over v}
+i{\p\log\M^2\over\p r}\right)+{\delta K_R\over vr^2}
\right\rbrace.\nonumber\\
\label{canonic}
\end{eqnarray}
With this new variable, the sonic point $\rso$ corresponds to 
$X\to +\infty$, while the spatial infinity corresponds to $X=0$.

\section{Approximations of the homogeneous solution}

\subsection{WKB approximation far from the accretor\label{Awkb}}

The general solution $f$ of the homogeneous equation
associated to Eq.~(\ref{canonic}) is a linear 
combination of ingoing ($f_+$) and outgoing  ($f_-$) waves, which can 
be approximated by the WKB method:
\begin{equation}
{\tilde f^\pm}\sim {\omega^{1\over2}\over W^{1\over4}}
\exp \left(\pm i\int^r {v W^{1\over2}\over 1-\M^2}\d r\right).
\label{defsonadiab}
\end{equation}
This approximation is asymptotically valid for
\begin{equation}
{\p \log W\over\p X}\ll W^{1\over2},\label{wkbhigh}
\end{equation}
which is satisfied at high frequency or far from the accretor.
The Wronskien ${\cal W}$ associated to the couple 
(${\tilde f^+},{\tilde f^-}$) of solutions is:
\begin{eqnarray}
{\cal W}&\equiv&{\tilde f^+}{\p {\tilde f^-}\over\p r}-
{\tilde f^-}{\p{\tilde f^+}\over\p r}, \\
&=&-{2i\omega v\over1-\M^2}.
\end{eqnarray}
The pressure perturbation $\Delta p$ associated to a perturbation $f$ 
is deduced from Eq.~(\ref{dber}):
\begin{equation}
{\Delta p\over\gamma p}={1\over c^2}
\e^{-i\omega\int^r{\M^2\over 1-\M^2}
{\d r\over v}}\left({{\tilde f}\over 1-\M^2}+{iv\over\omega}
{\p {\tilde f}\over \p r}\right).
\end{equation}
The normalization factor $\omega^{1\over2}$ in 
Eq.~(\ref{defsonadiab}) is such that the pressure perturbations 
$\Delta p_\pm$ associated to these solutions by Eq.~(\ref{dber}) 
have the following asymptotic behaviour:
\begin{equation}
{\Delta p^{\pm}\over \gamma p}\sim{c_\infty\over c}{\M^{1\over2}\over1\pm\M}
\e^{\mp i\omega {r\over c_{\infty}}}.\label{Deltap}
\end{equation}
Both carry the same acoustic flux $F^\pm={\dot M}_{0}c_\infty^{2}$ according 
to Eq.~(\ref{aflux}).

\subsection{Definition of the refraction coefficient 
$\bcmo$ and estimate at high frequency \label{Arefrac}}

\subsubsection{$\gamma<5/3$}

A Frobenius expansion in the vicinity of the sonic point ($\M=1$) 
shows that 
any solution of the homogeneous equation can be projected onto the 
couple $f_1,f_2$ of solutions, where $f_1(r)$ is regular and $f_2(r)$ 
is singular:
\begin{eqnarray}
f_1(r)&\sim& 1+{\cal O}(r),\label{frob1}\\
f_2(r)&\sim& \left\lbrack 1+{\cal O}(r)\right\rbrack
\exp \left\lbrace {i\omega\over 2\cs\dMs}
\log(r-\rso)\right\rbrace.\label{frob2}
\end{eqnarray}
This defines a unique coefficient $\bcmo$ such that the solution 
\begin{equation}
f_0 \equiv f^+ + \bcmo f^-,\label{fpm}
\end{equation}
is regular at the sonic point. The coefficient $|\bcmo|^{2}$ is the 
refraction coefficient of sound waves. The linearity of the equations 
guarantees of course that the solution $\Delta p_0 \equiv \Delta p^+ 
+ \bcmo \Delta p^-$ is also regular.\\
An expansion of $W$ in the vicinity of the sonic point leads to:
\begin{eqnarray}
W&=&q_0+q_1(r-\rso)+{\cal O}(r-\rso)^2,\\
q_0&\equiv&{\omega^2\over \cs^4},\\
q_1&\equiv&-{\omega^2\over \cs^4}
{\p\log c^2v^2\over\p r}+{2L^{2}\dMs\over\cs^2\rso^2},\\
r-\rso&\sim&\exp\left({2\dMs\over \cs}X\right).
\end{eqnarray}
The WKB approximation of ${\tilde f}^\pm$ is valid at high frequency 
$\omega\gg \omcut$, for $r-\rso\ll 1$:
\begin{equation}
{\tilde f}^\pm\sim \cs \exp\pm i\left\lbrack
{\omega\over 2\cs\dMs}\log(r-\rso)+{\cal O}(1)\right\rbrack.
\end{equation}
The asymptotic solution to the differential equation (\ref{homogene}) 
can also be approximated by a Bessel function in the vicinity of the sonic 
point, for $r-\rso\ll 1$. The regularity of $f_0$ at the sonic radius 
implies:
\begin{eqnarray}
{\tilde f_0} &\sim &\lambda J_\nu\left\lbrack 
{-\cs q_1^{1\over2}\over\dMs}(r-\rso)^{1\over2}\right\rbrack,
\label{f0}\\
\nu &\equiv& {i\cs q_0^{1\over2}\over\dMs}
={i\omega\over \cs\dMs}.
\end{eqnarray}
The Bessel and WKB approximations are both valid in the region where
$\omega^{-1}\ll r-\rso\ll 1$. In this region, the matching of the two 
approximations gives the normalization constant $\lambda$ and the 
refraction coefficient $\bcmo$. The asymptotic behavior of the Bessel 
function, when both the imaginary order $\nu$ and the argument 
tend to infinity, is obtained from Watson (1952, Chap.~8.6, p.262).
The Bessel function $J_{\nu}(z)$ is written in terms of the Hankel 
functions $H_{\nu}^{(1)}(z)$ and $H_{\nu}^{(2)}(z)$.
Let $x,z$ be real numbers, with $|x|,z\gg1$, such that the order
$\nu$ is purely imaginary $\nu\equiv iz\sinh x$:
\begin{eqnarray}
J_{\nu}(z)&=&{1\over2}\left\lbrack H_{\nu}^{(1)}(z)+
H_{\nu}^{(2)}(z)\right\rbrack,\\
H_{\nu}^{(1)}(z)&\sim&
\left\lbrack{2\exp(\pi z\sinh x)\over\pi z\cosh x}\right\rbrack^{1\over2}
\e^{+i\left\lbrack z(\cosh x-x\sinh 
x)-{\pi\over4}\right\rbrack},\label{hp}\\
H_{\nu}^{(2)}(z)&\sim&
\left\lbrack{2\exp(-\pi z\sinh x)\over\pi z\cosh x}\right\rbrack^{1\over2}
\e^{-i\left\lbrack z(\cosh x-x\sinh 
x)-{\pi\over4}\right\rbrack}.\label{hm}
\end{eqnarray}
Applying these asymptotic expansions to Eq.~(\ref{f0}), with
\begin{eqnarray}
z&\equiv&-{\cs q_1^{1\over2}\over\dMs}(r-\rso)^{1\over2}>0,\\
\sinh x &=& {\nu\over iz}\sim 
-\left|(r-\rso){\p\log c^{2}v^{2}\over\p r}\right|^{-{1\over2}}<0,
\end{eqnarray}
we obtain $\lambda$ and $|f_0(\rso)|$ to first order at high frequency:
\begin{eqnarray}
\lambda&= &\left|{2\pi\cs\omega\over\dMs}\right|^{1\over2}
\e^{\omega\pi\over 2\cs\dMs},\label{applambda}\\
|f_0(\rso)|&=&\cs.
\end{eqnarray}
The exponential decrease of $|\bcmo|$ at high frequency is deduced from 
Eqs. (\ref{hp}) and (\ref{hm}), by identifying the Hankel functions 
with the ingoing and outgoing acoustic waves:
\begin{equation}
|\bcmo|\propto \e^{\omega\pi\over \cs\dMs}.\label{expdecrease}
\end{equation}

\subsubsection{$\gamma=5/3$, $l=0$}

The WKB approximation is valid in the range $r\gg 1/\omega$.
\begin{equation}
\tilde f^\pm \sim c\M^{1\over2}\exp\pm i\omega\int_0^r {\M\over 
v(1-\M^2)} \d r,
\end{equation}
where the normalization is chosen in agreement with 
Eq.~(\ref{defsonadiab}).
In the region $r\le 1/\omega$ where the WKB approximation is not 
valid, we approximate the solution by the Bessel function $J_0$, which 
is regular at the origin:
\begin{eqnarray}
{\p^2\tilde f_0\over\p r^2} &+& {1\over r}\left\lbrack
1+{\cal O}(r^{1\over2})\right\rbrack{\p\tilde f_0\over\p r} + 
{\omega^2\over8}\tilde f_0\left\lbrack
1+{\cal O}(r^{1\over2})\right\rbrack = 0,\\
\tilde f_0&\sim &\tilde f_0(0)J_0\left({\omega r\over 2^{3\over2} 
}\right).
\end{eqnarray}
The other Bessel solution, $Y_{0}$, is singular with a logarithmic 
divergence, as in the case of $\gamma<5/3$.
The Bessel approximation for $r\gg1/\omega$ matches the WKB approximation:
\begin{eqnarray}
\tilde f_0(0)J_0\left({\omega r\over 2^{3\over2} }\right)&\sim &
\tilde f_0(0)\left({2^{5\over2}\over\pi \omega r}\right)^{1\over2}
\cos\left({\omega r\over 2^{3\over2} } - {\pi\over4}\right),\\
&\sim &
{1\over 2^{1\over2}\gamma r^{1\over2}}\left(\e^{-i{\omega r\over 
2^{3\over2}}}
+\bcmo\e^{+i{\omega r\over 2^{3\over2}}}\right).
\end{eqnarray}
From this we deduce that:
\begin{eqnarray}
\tilde f_0(0) &=& {\cal O}\left(\omega^{1\over2}\right),\\
|\bcmo | &=& 1.
\end{eqnarray}

\subsubsection{$\gamma=5/3$, $l\ge1$}

The case of non radial perturbations needs to be treated separately,
because the behaviour of the solution is quite different near the 
sonic point. If $l\ge1$, there is always a turning point beyond which 
the solution is evanescent, such that it decays exponentially on the 
sonic point. The leading behaviour of $f_{0},g_{0}$ on this essential 
singularity is the following, within a multiplicative constant:
\begin{eqnarray}
f_{0}&\sim& r^{1\over 8}\exp-{2L\over r^{1\over4}},\label{f530}\\
g_{0}&\sim&2^{3\over2}{iL\over\omega r^{1\over8}}\exp-{2L\over r^{1\over4}}.
\label{g530}
\end{eqnarray}
The solutions of the other branch diverge exponentially on the sonic 
point: as in the case of a regular sonic point, there is a unique 
combination of ingoing and outgoing waves such that the solution does 
not diverge at the sonic point.
The turning point $r_{\rm t}$ corresponds to the radius where $W=0$ 
in Eq.~(\ref{canonic}): 
\begin{equation}
r_{\rm t}\sim  \left({2L^2\over\omega^2}\right)^{2\over5}.\label{turn}
\end{equation}
The WKB approximation is valid for $r_{\rm t}\ll r\ll 1$:
\begin{eqnarray}
\tilde f_\pm&\sim &{c\M^{1\over2}\over 
\left\lbrack 1-{L^2c^2\over 
r^2\omega^2}(1-\M^2)\right\rbrack^{1\over4}}
\times \nonumber\\
& &\exp\pm i\omega\int_{r_0}^{r} {\M\over v(1-\M^2)}
\left\lbrack 1-{L^2c^2\over 
r^2\omega^2}(1-\M^2)\right\rbrack^{1\over2}
\d r.\label{wkbl}
\end{eqnarray}

\section{Acoustic efficiencies $\cpl,\vpl$ of entropy and vorticity 
perturbations}

\subsection{Acoustic efficiency $\vpl$}

The outgoing pressure perturbation produced by the advection of a  
perturbation $\delta K_R$ is computed by solving 
Eq.~(\ref{canonic}), imposing the regularity at the sonic point and the 
absence of an incoming sound wave from infinity. The general solution of 
Eq.~(\ref{canonic}) with $\delta S_R=0$ can be written as:
\begin{eqnarray}
{\tilde f}&=&-{i\delta K_R\over2\omega}\left\lbrack {\tilde f^-}
\left\lbrace\beta +\int_{r_0}^r{{\tilde f^+}\over r^2v}
\e^{i\omega\int^r{\d r\over v(1-\M^2)}}\right\rbrace
-\right.\nonumber\\
&&\left.{\tilde f^+}
\left\lbrace\alpha +\int_{r_0}^r{{\tilde f^-}\over r^2v}
\e^{i\omega\int^r{\d r\over v(1-\M^2)}}\right\rbrace\right\rbrack,
\label{solgen}\\
{\delta p\over\gamma p}&=&-{i\delta K_R\over2\omega}\left\lbrack 
{\Delta p^-\over\gamma p}
\left\lbrace\beta +\int_{r_0}^r{{\tilde f^+}\over r^2v}
\e^{i\omega\int^r{\d r\over v(1-\M^2)}}\right\rbrace
-\right.\nonumber\\
&&\left.{\Delta p^+\over\gamma p}
\left\lbrace\alpha +\int_{r_0}^r{{\tilde f^-}\over r^2v}
\e^{i\omega\int^r{\d r\over v(1-\M^2)}}\right\rbrace\right\rbrack.
\label{solgenpv}
\end{eqnarray}
According to Eqs.~(\ref{Deltap}) and (\ref{frob1}-\ref{frob2}), the 
integrals involved in Eq.~(\ref{solgenpv}) are well defined both at 
infinity and at the sonic radius. The condition that no sound wave 
comes from infinity leads to:
\begin{equation}
\alpha = -\int_{r_0}^{\infty}{{\tilde f^-}\over r^2v}
\e^{i\omega\int^r{\d r\over v(1-\M^2)}}.
\end{equation}
The regularity at the sonic radius can be written:
\begin{equation}
\beta = -\bcmo \alpha-\int_{r_0}^{\rso}{{\tilde f_0}\over r^2v}
\e^{i\omega\int^r{\d r\over v(1-\M^2)}} .
\end{equation}
Thus the efficiency of sound wave emission by the advection of 
vorticity perturbations is:
\begin{equation}
\vpl = -{i\over2\omega}\int_{\rso}^\infty
\e^{i\omega\int_R^r {1+\M^2\over1-\M^2}{\d r\over v}}
{f_0\over r^2 v}\d r.\label{conversionv}
\end{equation}
This integral converges as $1/r^{3}$ at infinity. It can be transformed 
using integrations by parts together with the homogeneous differential 
system satisfied by $f_{0},g_{0}$, so that it converges rapidly at infinity, 
thus allowing an easier numerical computation. The following expression
converges as $1/r^{5}$:
\begin{eqnarray}
\vpl = {1\over2\omega^{2}}&&\int_{\rso}^\infty
\e^{i\omega\int_R^r {1+\M^2\over1-\M^2}{\d r\over v}}\nonumber\\
&&\times\left\lbrace
{\p\over\p r}\left({1-\M^{2}\over r^{2}}\right)f_0
+{i\omega v\over r^{2}}g_{0}
\right\rbrace\d r.\label{fastqk}
\end{eqnarray}

\subsection{Acoustic efficiency $\cpl$}

The outgoing pressure perturbation produced by an incoming entropy 
perturbation $\delta S_R$ is also computed by solving 
Eq.~(\ref{canonic}), imposing the regularity at the sonic point and 
the absence of an incoming sound wave from infinity. The general 
solution of Eq.~(\ref{canonic}) with $\delta K_R=0$ can be written as:
\begin{eqnarray}
{\tilde f}=-{\delta S_R\over2\gamma}\times\nonumber\\
\left\lbrack {\tilde f^-}
\left\lbrace\beta +\int_{r_0}^r{\tilde f^{+}}{\p\over\p r}
\left( {1-\M^2\over \M^2}\e^{i\omega\int^r{\d r\over v(1-\M^2)}}
\right)\right\rbrace
-\right.\nonumber\\
\left.{\tilde f^{+}}
\left\lbrace\alpha +\int_{r_0}^r{\tilde f^-}{\p\over\p r}
\left( {1-\M^2\over \M^2}\e^{i\omega\int^r{\d r\over v(1-\M^2)}}
\right)\right\rbrace\right\rbrack,
\label{solgene}\\
{\delta p\over\gamma p}=-{\delta S_R\over2\gamma}\times\nonumber\\
\left\lbrack 
{\Delta p^-\over\gamma p}
\left\lbrace\beta +\int_{r_0}^r{\tilde f^{+}}{\p\over\p r}
\left( {1-\M^2\over \M^2}\e^{i\omega\int^r{\d r\over v(1-\M^2)}}
\right)\right\rbrace
-\right.\nonumber\\
\left.{\Delta p^{+}\over\gamma p}
\left\lbrace\alpha +\int_{r_0}^r{\tilde f^-}{\p\over\p r}
\left( {1-\M^2\over \M^2}\e^{i\omega\int^r{\d r\over v(1-\M^2)}}
\right)\right\rbrace
\right\rbrack.
\label{solgenpe}
\end{eqnarray}
Since the integrals of Eq.~(\ref{solgenpe}) converge at the sonic 
point, the regularity of the solution at the sonic point can be 
written:
\begin{equation}
\beta = -\alpha\bcmo-\int_{r_0}^r{\tilde f_0}{\p\over\p r}
\left( {1-\M^2\over \M^2}\e^{i\omega\int^r{\d r\over v(1-\M^2)}}
\right).\label{beta}
\end{equation}
In Eq.~(\ref{solgenpe}) the integrals diverge at infinity. Two 
integrations by parts leads to define the two functions $F_\pm, H_\pm$ 
as follows:
\begin{eqnarray}
F^\pm &\equiv &c^2{(1-\M^2)^2\over\M^2}{\Delta p^\pm\over\gamma p}
\e^{i\omega\int^r{1+\M^2\over1-\M^2}{\d r\over v}},\\
H^\pm &\equiv&\e^{i\omega\int^r{\d r\over v(1-\M^2)}}{\p\over\p 
r}\left\lbrack{v(1-\M^2)^2\over i\omega\M^2}
{\p\tilde f^\pm\over\p r}\right\rbrack.
\end{eqnarray}
The pressure perturbation associated to the entropy perturbation 
$\delta S_R$ is: 
\begin{eqnarray}
{\delta p\over \gamma p}&=& -{\delta S_R\over2\gamma}\left\lbrack
{\Delta p^-\over \gamma p }
\left\lbrace \beta + \left\lbrack F^{+}(r_0) + 
\int_{r_0}^r H^{+}\d r\right\rbrack\right\rbrace\right.\nonumber\\
&+& \left.{\Delta p^{+}\over \gamma p }
\left\lbrace \alpha - \left\lbrack F^-(r_0) + 
\int_{r_0}^r H^-\d r\right\rbrack\right\rbrace\right\rbrack.\label{dp}
\end{eqnarray}
Far from the accretor, the pressure perturbation can be projected 
onto the ingoing and outgoing sound waves. $\alpha$ is chosen such 
that there is no sound wave coming from infinity:
\begin{equation}
\alpha =  F^-(r_0)+\int_{r_0}^\infty H^-\d r.
\label{alpha}
\end{equation}
Using Eqs.~(\ref{beta}) and (\ref{alpha}), and the fact that 
$F_0(\rso)=0$, we take the limit $r\to\infty$,  $r_0\to \rso$ in 
Eq.~(\ref{dp}) and obtain:
\begin{equation}
\cpl = {i\over 2\gamma\omega}\int_{\rso}^\infty
\e^{i\omega\int^r{\d r\over v(1-\M^2)}}
{\p\over\p r}\left\lbrack{v(1-\M^2)^2\over\M^2}
{\p \tilde f_0\over\p r}\right\rbrack\d r.
\label{conversion}
\end{equation}
This integral can be written with the following format, using the 
homogeneous differential system (Eqs.~\ref{dtfdr}, \ref{dtgdr}) 
satisfied by $f_{0},g_{0}$:
\begin{equation}
\cpl ={1\over 2\gamma}\int_{\rso}^\infty
\e^{i\omega\int^r{1+\M^{2}\over1-\M^{2}}{\d r\over v}}
(A_{k} f_{0}+B_{k}g_{0})\d r,
\end{equation}
where the functions $A_{k}(r),B_{k}(r)$ depend only on the unperturbed flow,
and the integral converges as $1/r^k$ far from the accretor:
\begin{eqnarray}
A_{1} &\equiv& - {i\omega\over v}
\left(1-{\omega_{l}^2\over\omega^{2}}\right),\\
B_{1}&\equiv& -{\p\over\p r}(c^{2}-v^2).
\end{eqnarray}
Two successive integrations by parts make the convergence of the 
integral much faster, thus allowing an easier numerical calculation:
\begin{eqnarray}
A_{5}&\equiv&{\p\over\p r}\left\lbrack
(1-\M^{2})\left(1-{\omega_{L}^{2}\over\omega^{2}}\right)
\right\rbrack\nonumber
\\
&+&{1\over c^{2}}\left(1-{\omega_{L}^{2}\over\omega^{2}}\right)
{\p\over\p r}(c^{2}-v^2)-
{i\omega v\over 
c^{2}}\left(1-{\omega_{L}^{2}\over\omega^{2}}\right)^2,\\
B_{5}&\equiv& {\p\over\p r}\left\lbrace (1-\M^{2})\left\lbrack
{v\over i\omega}{\p\over\p r}(c^{2}-v^2)
-v^{2}\left(1-{\omega_{L}^{2}\over\omega^{2}}\right)\right\rbrack
\right\rbrace.\label{fastinteg}
\end{eqnarray}

\section{Asymptotic estimate of $|\cpl|,|\vpl|$ at high frequency 
for $\gamma=5/3$\label{app53}}

Eq.(\ref{conversion}) can be estimated in the limit $\omega\gg 
c_\infty^3/GM$, when $\gamma=5/3$. We treat the case of 
radial ($l=0$) and non radial ($l\ge1$) perturbations separately.

\subsection{Radial perturbations $l=0$\label{appradial}}

We use a local expansion of $\M$ where $r\ll 1$:
\begin{equation}
\M^2 = 1 - 4r^{1\over2} + {\cal O}(r).
\end{equation} 
Writing $\tilde f_0=\tilde f_- + \bcmo\tilde f_+$, we obtain
for $r_1\gg1/\omega $ and $r_2\ll1$:
\begin{eqnarray}
\int_{r_1}^{r_2}\e^{i\omega\int_R^r{\d r\over v(1-\M^2)}}
{\p\over\p r}\left({v(1-\M^2)^2\over\M^2}{\p \tilde f_0\over\p 
r}\right)\d r
\sim \nonumber\\
{6^{2\over3}\over 5}\omega^{1\over3}\bcmo
\int_{u_1}^{u_2}{\e^{-iu}\over u^{1\over3}}\d u,\nonumber\\
u\equiv {2^{1\over2}r^{3\over2}\omega\over 3}.
\end{eqnarray}
The range of validity  $1/\omega\ll r\ll 1$ implies that $u_1\gg 
1/\omega^{1/2}$ and $u_2\ll\omega$. The contribution of the region 
$r_2\le r\le \infty$ is negligible.
\begin{equation}
\int_{r_1}^{r_2}\e^{i\omega\int_R^r{\d r\over v(1-\M^2)}}
{\p\over\p r}\left({v(1-\M^2)^2\over\M^2}{\p \tilde f_0\over\p 
r}\right)\d r
\ll \omega^{1\over 3}.
\end{equation}
The contribution of the region $r\le r_1$ is estimated using the 
asymptotic behaviour of the Bessel function $J_1$:
\begin{eqnarray}
-{\p\over\p z}\left({1-\M^2\over\M^2}{\p\tilde f_0\over\p z}\right)
\sim 2^{11\over4} {\tilde f_0(0)\over\omega^{1\over2}}
{\p\over\p z}\left\lbrack z^{1\over2}J_1(z)\right\rbrack,\\
\int_{0}^{z_1}\e^{-iz}
{\p\over\p z}\left({1-\M^2\over\M^2}{\p \tilde f\over\p z}\right)\d z
\sim {\tilde f_0(0)\over\omega^{1\over2}}
{2^{9\over4}\over\pi^{1\over2}}\e^{-i{\pi\over4}}z_1
\end{eqnarray}
Thus we deduce the acoustic efficiency at high frequency:
\begin{equation}
\cpl\sim -i\e^{iz_R}{\bcmo\over\gamma}
\left({\omega\over6}\right)^{1\over3}
\int_0^\infty{\e^{-iu}\over u^{1\over3}}\d u.\label{cplu}
\end{equation}
The main contribution to the integral comes from the region where 
$u\sim 1$, \ie $r_{\rm eff}\sim (6/\omega)^{2/3}/2$. 
Eq.~ (\ref{cplu}) is rewritten in Eq.~(\ref{abscpl}) using the Gamma 
function.

\subsection{Non radial perturbations $l\ge1$\label{appm}}

As in the case of radial perturbations, the main contribution to the 
integral comes from the region
accessible to the WKB approximation.
The outgoing wave is the main contributor to the integral 
(\ref{conversion}). Using the asymptotic development of $v,c$ for 
$r\ll1$ in Appendix.~\ref{Aadiab},
\begin{eqnarray}
\int_{r_0}^{r} \left\lbrace 1-\M
\left\lbrack 1-{L^2c^2\over 
r^2\omega^2}(1-\M^2)\right\rbrack^{1\over2}
\right\rbrace{\d r\over v(1-\M^2)}
\sim \nonumber\\
{1\over3}\left(-2^{1\over2}r^{3\over2}\omega
+{L^2\over 2^{1\over2}r^{3\over2}\omega}\right)\label{apint}.
\end{eqnarray}
With $u\equiv 2^{1\over2}r^{3\over2}\omega/3$, the acoustic 
efficiencies at high frequency are deduced from Eqs.~(\ref{wkbl}), 
(\ref{conversionv}), (\ref{conversion}) and (\ref{apint}):
\begin{eqnarray}
\cpl &\sim &i{\bcmo\over\gamma}
\left({\omega\over6}\right)^{1\over3}
\int_0^\infty \exp-i\left(u-{L^2\over9u}\right){\d u\over 
u^{1\over3}},\label{cpli}\\
\vpl &\sim & i{2^{1\over3}\bcmo\over3^{5\over3}\omega^{1\over3}}
\int_0^\infty \exp-i\left(u-{L^2\over9u}\right){\d u\over 
u^{5\over3}}.
\label{vpli}
\end{eqnarray}
Using the new variable
\begin{equation}
t\equiv {3\over 2L}\left\lbrack u - {L^2\over9u}\right\rbrack,
\end{equation}
the integrals in Eqs.~(\ref{cpli}-\ref{vpli}) can be written using 
$K_{2\over3}$, the modified Bessel function of order $2/3$ 
(Gradshteyn \& Ryzhik 1980, p. 430):
\begin{eqnarray}
\int_0^\infty\exp-i\left(u-{L^2\over9u}\right){\d u\over u^{1\over3}}
&=&2\left({L\over3}\right)^{2\over3}\e^{-i{\pi\over3}}K_{2\over3}
\left({2L\over3}\right),\\
\int_0^\infty\exp-i\left(u-{L^2\over9u}\right){\d u\over u^{5\over3}}
&=&2\left({3\over L}\right)^{2\over3}\e^{+i{\pi\over3}}K_{2\over3}
\left({2L\over3}\right).
\end{eqnarray}
Thus we obtain:
\begin{eqnarray}
\cpl &\sim &{2i\over3\gamma}
\bcmo \left({\omega L^2\over2}\right)^{1\over3}\e^{-{i\pi\over3}}
K_{2\over3}\left({2L\over3}\right),\\
\vpl &\sim & {2i\over3}\bcmo\left({2\over\omega 
L^2}\right)^{1\over3}\e^{+{i\pi\over3}}
K_{2\over3}\left({2L\over3}\right).
\end{eqnarray}
Here again, the main contribution to the integrals comes from the 
region $r_{\rm eff}\sim (6/\omega)^{2/3}/2$.
By contrast with $|\cpl|$, $|\vpl|$ decreases at high frequency.

\end{document}